\documentclass[aps,twocolumn,superscriptaddress,amsmath]{revtex4-1}

\pdfoutput = 1

\usepackage[pdftex]{graphicx,color}
\usepackage{soul}
\usepackage{amsfonts}
\usepackage{amssymb}
\usepackage{amsmath}
\usepackage{amsthm}
\usepackage{ulem}

\usepackage{subfigure}
\usepackage{rotate}
\usepackage{bm}
\usepackage{dcolumn}

\usepackage[pdftex]{hyperref}

\let\oldmarginpar\marginpar
\renewcommand\marginpar[1]{\-\oldmarginpar[\raggedleft\tiny #1]
{\raggedright\tiny #1}}

\newcommand{\avg}[1]{\left< #1 \right>}
\newcommand{\bra}[1]{\langle#1|}
\newcommand{\ket}[1]{|#1\rangle}
\newcommand{\braket}[2]{\langle#1|#2\rangle}

\graphicspath{{figs/}}

\begin{document}

\title{The many-body localized phase of the quantum random energy model}

\author{C. L. Baldwin}
\affiliation{Department of Physics, University of Washington, Seattle, WA 98195, USA}

\author{C. R. Laumann}
\affiliation{Department of Physics, University of Washington, Seattle, WA 98195, USA}

\author{A. Pal}
\affiliation{Department of Physics, Harvard University, Cambridge, MA 02138, USA}
\affiliation{Rudolf Peierls Centre for Theoretical Physics, Oxford University, Oxford OX1 3NP, UK}

\author{A. Scardicchio}
\affiliation{Abdus Salam ICTP Trieste, Strada Costiera 11, 34151 Trieste, Italy}
\affiliation{INFN, Sezione di Trieste, Via Valerio 2, 34127 Trieste, Italy}
\affiliation{Dipartimento di Fisica, Universit\`a degli Studi di Bari ``Aldo Moro'', I-70126, Bari, Italy}
\date{\today}

\begin{abstract}

The random energy model (REM) provides a solvable mean-field description of the equilibrium spin glass transition. Its quantum sibling (the QREM), obtained by adding a transverse field to the REM, has similar properties and shows a spin glass phase for sufficiently small transverse field and temperature. In a recent work, some of us have shown that the QREM further exhibits a many-body localization - delocalization (MBLD) transition when viewed as a closed quantum system, evolving according to the quantum dynamics. This phase encloses the familiar equilibrium spin-glass phase. In this paper we study in detail the MBLD transition within the forward-scattering approximation and replica techniques. The predictions for the transition line are in good agreement with the exact diagonalization numerics. We also observe that the structure of the eigenstates at the MBLD critical point changes continuously with the energy density, raising the possibility of a family of critical theories for the MBLD transition.
\end{abstract}

\maketitle

\tableofcontents

\section{Introduction} \label{sec:introduction}

Experiments on few-atom systems show striking contradictions with the extension of classical laws to arbitrarily small distances and energies. Quantum mechanics has proven to be a successful theory for the dynamics of these microscopic systems. Further developments make clear that quantum effects are also important for macroscopic systems, although typically at low temperature or high density.

On the other hand, at sufficiently high temperature or low density, the semiclassical principle appears to state that quantum and classical dynamics yield the same results for most measurable quantities. It is then interesting to examine the cases in which this reconciliation does not occur. Disordered systems make a particularly interesting playground. Non-interacting particles in a disordered potential can exhibit Anderson localization \cite{anderson1958absence}: the gas' diffusion coefficient(s) are completely suppressed, even in a regime where the classical dynamics remains diffusive. This difference is so spectacular that in low dimensions the dynamics at all energy scales supported by the system is localized. Here the classical description is never, not even qualitatively, accurate.

Whether Anderson localization survives the introduction of interactions between particles or not has been a subject of debate since the beginning of the field \cite{mott1969conduction,fleishman1980interactions}. Yet in the last decade, the work generated by the seminal paper of Basko, Aleiner and Altshuler \cite{basko2006metal} on interacting, disordered systems has shed considerable light on the issue \cite{oganesyan2007localization,znidaric2008many,pal2010mb}.

It is now clear that a sufficiently small interaction is for many purposes irrelevant (one exception being entanglement \cite{Bardason2012,nanduri2014entanglement}). The system behaves as if the single-particle occupation numbers are ``perturbatively dressed" into the interacting phase \cite{huse2014phenomenology,serbyn2013local,imbrie2014many,ros2015integrals,chandran2015constructing}. 
Excitations in the ``many-body localized'' (MBL) regime are localized, transport is suppressed, and the dynamics is not thermalizing. For some spin chains, researchers have even been able to pinpoint a transition between an MBL region and an ergodic region \cite{pal2010mb,de2013ergodicity,kjall2014many,luitz2015many,Mondragon-Shem:2015aa,goold2015total}. The properties of this transition are not well understood. A few elementary constraints have been imposed \cite{Grover:2014aa,Chandran:2015ab} and similarities with infinite randomness fixed points have been found in one-dimensional studies \cite{pal2010mb,vosk2013many,Potter:2015ab}. Like the Anderson transition, this is not a usual thermodynamic phase transition but rather a \emph{dynamical} phase transition. There is no local order parameter in terms of which to write a Landau-Ginzburg free energy density. The transition itself is the breakdown of the hypothesis under which one derives the statistical description from the underlying microscopic equations of motion. Consequently, this transition can be observed even at infinite temperature.

An even more recent line of research studies how ergodicity breaking in the quantum dynamics compares to that of more canonical (classical and quantum) glassy phases. Since MBL is easily observed in spin chains with quenched disorder (and also a phase more akin to configurational glasses has been conjectured \cite{schiulaz2013ideal,pino2015non,Yao:2014aa,Papic:2015aa}), it is natural to look for MBL in spin glasses.

With this in mind, some of us have recently looked \cite{laumann2014many} at a quantum version of Derrida's random energy model (REM) \cite{derrida1981random}, which is a simplified model of mean-field spin glass. The quantum model's equilibrium phase diagram has been studied before \cite{goldschmidt1990solvable} and a glassy phase was identified at low temperature and small transverse magnetic field \footnote{A similar phase diagram is found in more realistic, still mean-field quantum spin glasses \cite{laumann2008cavity}. This should be generic for a large family of models, including combinatorial optimization problems in quantum annealers \cite{laumann2015quantum}.}.

In \cite{laumann2014many} it was found that the quantum dynamics is ergodic for high temperatures and large transverse field, but ergodicity breaks down upon lowering the temperature and the transverse field. The ergodicity-broken phase, which we identify as the MBL phase, encompasses the glassy region. Therefore the quantum dynamics becomes non-ergodic before the glassy phase sets in, disentangling the concept of ergodicity breaking from that of replica symmetry breaking in these quantum models. A qualitatively similar observation was recently made by studying Rokhsar-Kivelson-type wavefunctions derived from the REM~\cite{chen2015manybody}. However, on further thought, what is really surprising is that there is an ergodic phase for the simple transverse-field quantum dynamics, since the classical Monte Carlo dynamics of the REM is \emph{never} ergodic. This observation is relevant for the science of quantum annealers, for which the MBL phase could be a significant stumbling stone \cite{altshuler2010anderson,laumann2015quantum}. 

In this paper we analyze in much more detail the MBL phase of the QREM and the transition between the ergodic and MBL phases. We also study in detail the application of the forward-scattering approximation (FSA) in the MBL phase. We find that the localized phase is consistently distinct from the ergodic phase in its level-spacing statistics, observables, and eigenstate structure. Naive perturbation theory cannot accurately characterize the MBL phase, but by carefully handling near-degeneracies in the FSA, we quantitatively describe both the localized eigenstates and the phase boundary within perturbation theory. We accomplish this using a combination of simple approximations, numerics and a replica treatment of the forward-scattering wavefunctions.

The QREM is the first model in which a many-body mobility edge was clearly observed in the numerics  accompanied by an analytical prediction within the forward-scattering approximation. It is an ideal test bed to discuss the properties of the mobility edge. One of the things that we observe numerically is that the critical statistics of the eigenvalues changes continuously along the mobility edge. It is worth mentioning the analogy with the critical properties of mean-field spin glasses \cite{sompolinsky82,parisi13}, which also change continuously with lowering the temperature.

As a final remark, we notice that on lowering the transverse field the mobility edge shifts to higher temperatures. Furthermore, a mobility edge opens up even for an infinitesimal transverse field, in contrast to the situation in one-dimensional systems, where the MBL phase at infinite temperature is stable upto a finite value of the interaction. This is most probably a special feature of the infinite dimensionality of the QREM.

\section{Phenomena} \label{sec:phenomena}

The quantum random-energy model (QREM) for $N$ spin-$1/2$s is defined by
\begin{equation} \label{eq:Hamiltonian}
H = H_0(\{\hat{\sigma}_i^z\}) - \Gamma \sum_{i=1}^N \hat{\sigma}_i^x.
\end{equation}
Here, $\Gamma$ is the transverse field, and $H_0(\{\hat{\sigma}_i^z\})$ is a random operator diagonal in the $\{\hat{\sigma}_i^z\}$ basis, with the diagonal entries identically and independently distributed according to
\begin{equation} \label{eq:distribution}
P(E_0) = \frac{1}{\sqrt{\pi N}} e^{-\frac{E_0^2}{N}}.
\end{equation}
With this normalization the spectrum of $H_0$ is with high probability contained in $[-N\sqrt{\ln 2},+N\sqrt{\ln 2}]$. We note that throughout the paper, capitalized $E$ represents extensive energy and $\epsilon = E/N$ the corresponding energy density.

\subsection{Equilibrium Phase Diagram} \label{subsec:equilibrium_phase_diagram}

Goldschmidt~\cite{goldschmidt1990solvable} determined the canonical phase diagram of the QREM by using the Suzuki-Trotter expansion and replica trick. 
The results we give here are taken from his paper, in which detailed derivations are found as well. 
See also J\"{o}rg et al.~\cite{jorg2008simple} for a simple perturbative derivation of some parts of the phase diagram.

The QREM has three equilibrium phases:
\begin{itemize}
	\item The REM paramagnet. The free energy density is
		\begin{equation}
		f_{\textrm{REM}} = -T \ln{2} - \frac{1}{4T}.
		\end{equation}
		This is equal to the free energy density of the classical REM at temperature above 
		\begin{equation} \label{eq:classical_T_c}
		T_c \equiv \frac{1}{2\sqrt{\ln{2}}}.
		\end{equation}
		All thermodynamic quantities are identical to the zero-field REM, in particular the energy density $\epsilon = -\frac{1}{2T}$ and the entropy density $s = \ln{2} - \epsilon ^2$.
	\item The quantum paramagnet. The free energy density is
		\begin{equation}
		f_{\textrm{Q}} = -T \ln{2} - T \ln{\left( \cosh{\frac{\Gamma}{T}} \right) }.
		\end{equation}
		Note that this is the free energy density for non-interacting spins in a field $\Gamma$. The REM term in the Hamiltonian does not influence the equilibrium physics of this phase. Thermodynamic properties such as the energy, entropy, and magnetization density are given by the standard formulae.
	\item The spin-glass. The free energy density is simply
		\begin{equation}
		f_{\textrm{SG}} = -\sqrt{\ln{2}},
		\end{equation}
		identical to the classical REM. The system is frozen in its ground state at $\epsilon_0 = -\sqrt{\ln{2}}$, where there are $O(1)$ states and thus $s = 0$.
\end{itemize}

\begin{figure}[t]
\begin{center}
\includegraphics[width=1.0\columnwidth]{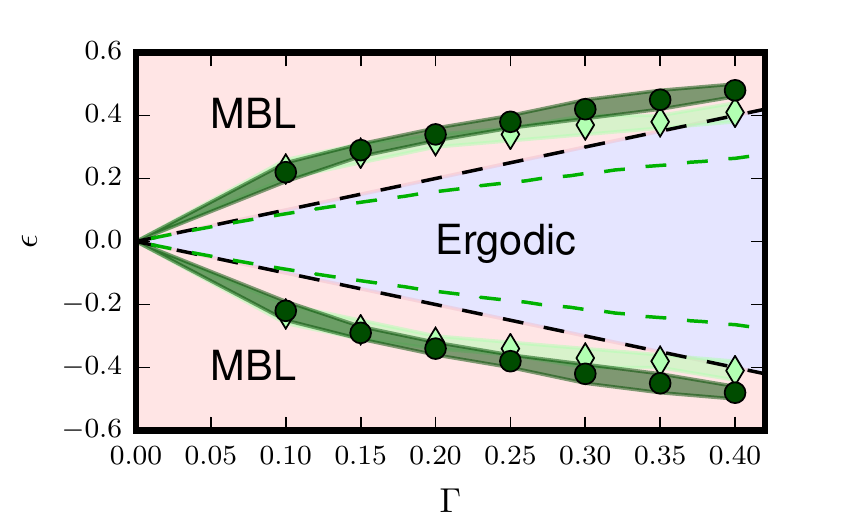}
\includegraphics[width=1.0\columnwidth]{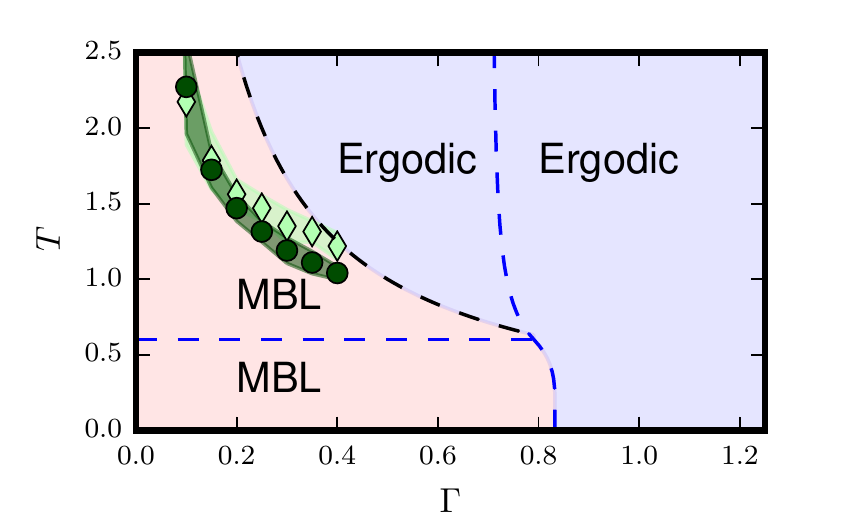}
\caption{The phase diagram of the QREM, in the $\Gamma - \epsilon$ plane (top) and the $\Gamma - T$ plane (bottom). The red shaded region contains localized eigenstates and the blue shaded region contains ergodic eigenstates. The green shaded regions indicate the numerically observed boundary region: dark green circles from exact diagonalization (Sec.~\ref{sec:exact_diagonalization}), and light green diamonds from the numerical FSA (Sec.~\ref{subsec:numerical_treatment}). The black dashed lines are the conjectured limiting boundaries of $\epsilon _c = \pm \Gamma$, the green dashed lines are the analytic estimate of $\epsilon_c$ within the single-resonance approximation (Sec.~\ref{subsec:single_resonance_approximation}), and (bottom only) the blue dashed lines indicate the thermodynamic phase boundaries as predicted by canonical calculations~\cite{goldschmidt1990solvable}.}
\label{fig:micro_phase}
\end{center}
\end{figure}

The transition lines between these phases are located where the free energies are equal. The boundary between the REM paramagnet and the spin-glass is as in the classical model: the system is paramagnetic for $T > T_c$ and frozen in the spin-glass phase for $T < T_c$ ($T_c$ given by Eq.~\eqref{eq:classical_T_c}). The system undergoes a first-order transition to the quantum paramagnet when $\Gamma$ increases past $\Gamma_c$, where $\Gamma_c$ is defined for $T > T_c$ by $f_{\textrm{REM}} = f_{\textrm{Q}}$ and for $T < T_c$ by $f_{\textrm{SG}} = f_{\textrm{Q}}$. The blue dashed lines of Fig.~\ref{fig:micro_phase}b mark these boundaries. In particular, note that $\Gamma_c$ at $T = 0$ is the field strength at which the ground state energy of the quantum paramagnet matches the ground state of the classical REM.

The essence of the canonical phase diagram is that the system makes a first-order transition between ``REM'' physics and ``quantum paramagnet'' physics. REM physics results from an otherwise structureless system freezing into an intensive number of configurations at non-zero temperature. Quantum paramagnet physics is that of non-interacting spins in a magnetic field. The QREM does not compromise between these two regimes. It exhibits only one or the other, at least thermodynamically.

\subsection{Dynamical Phase Diagram} \label{subsec:dynamical_phase_diagram}

When treated as an isolated system, the QREM has two dynamical phases:

\begin{itemize}
	\item The ergodic phase. Eigenstates satisfy the Eigenstate Thermalization Hypothesis (ETH) \cite{srednicki1994chaos,rigol2012thermalization}: expectation values of local observables agree with the microcanonical ensemble, which, here, is paramagnetic. Thus $\avg{\hat{\sigma}_i^z} = 0$ for all spins $i$. Fluctuations in $\sigma_i^z$ are large but decay exponentially in time. In addition, these eigenstates are delocalized in the following sense: one can map a configuration of $N$ spin-$1/2$'s to a corner of an $N$-dimensional hypercube by considering $\sigma_i^z = 1 (-1)$ as the top (bottom) face of the cube's $i$'th dimension. The QREM Hamiltonian is then an Anderson model on the corners of this hypercube, with the spin configurations being ``lattice sites''. Ergodic eigenstates are delocalized over these sites. The probability overlap between two ergodic states decays exponentially with their energy difference, as observed in the Anderson model \cite{evers08}.
	\item The many-body-localized phase. Eigenstates are weakly-dressed single configurations of spins. Thus $\avg{\hat{\sigma}_i^z} = \pm 1$. Fluctuations within an eigenstate are small (and decrease with system size), but fluctuations between eigenstates over energy and realization of disorder are order-$1$. These eigenstates are Anderson-localized on the hypercube. The probability overlap between a pair of eigenstates decays as $\omega ^{-2}$, where $\omega$ is the energy difference of the two states. This is well-described by first-order perturbation theory.
\end{itemize}

\begin{figure}[t]
\begin{center}
\includegraphics[width=1.0\columnwidth]{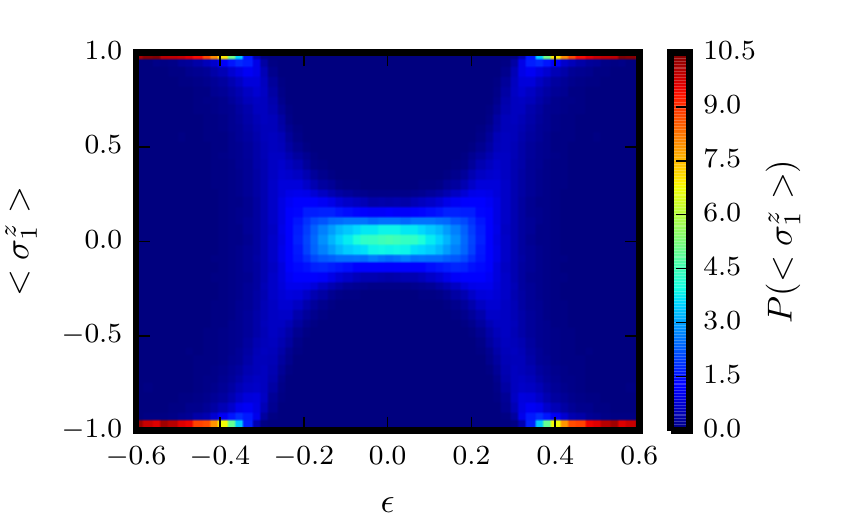}
\caption{Probability density function for the eigenstate single-spin magnetization (y-axis), for small bins over a range of energy densities (x-axis). Each vertical slice is a separate probability distribution. These distributions are at $\Gamma = 0.20$ and $N = 14$, with energy density windows $\delta \epsilon = 0.02$.}
\label{fig:ED_bug_diagram}
\end{center}
\end{figure}

\begin{figure}[t]
\begin{center}
\includegraphics[width=1.0\columnwidth]{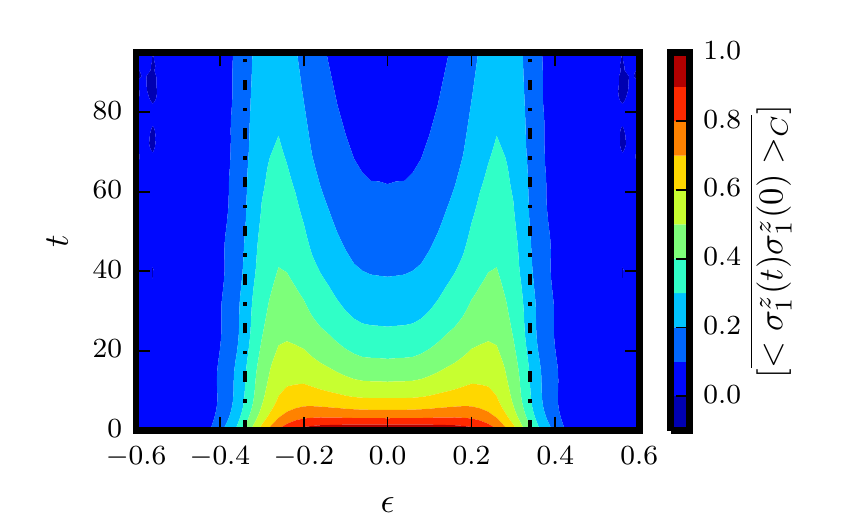}
\caption{Eigenstate autocorrelation function $<\sigma _1^z(t) \sigma _1^z(0)>_C$ (see Eq.~\eqref{eq:autocorrelation_definition}) as a function of time (vertical axis) and eigenstate energy density (horizontal axis). The vertical dashed lines indicate the location of the MBLD transition as determined by the QREM's spectral statistics. These results are taken at $\Gamma = 0.20$ and $N = 13$ with energy bins $\delta \epsilon = 0.02$.}
\label{fig:ED_time_evolution}
\end{center}
\end{figure}

A major focus of the present work is to quantitatively locate and characterize the boundary between these phases in the $\Gamma$-$\epsilon$ plane ($\epsilon$ is eigenstate energy density). We define order parameters with well-defined localized and ergodic limits for each characteristic described above. See Figs.~\ref{fig:ED_bug_diagram} and ~\ref{fig:ED_time_evolution} for examples and Sec.~\ref{sec:exact_diagonalization} for details. Since the states at $+\epsilon$ are statistically equivalent to those at $-\epsilon$, the phase boundary must be symmetric about $\epsilon = 0$. We focus on the negative-$\epsilon$ portion, denoted $\epsilon_c (\Gamma)$. Numerical evidence and perturbation theory each give constraints on $\epsilon_c (\Gamma)$. We show these, overlaid on our conjectured phase diagram, in Fig.~\ref{fig:micro_phase}.

Our numerical results indicate a transition between two phases and provide a lower bound for $\epsilon_c (\Gamma)$. The dark green markers and shading in Fig.~\ref{fig:micro_phase} indicate where order parameters computed via exact diagonalization (Sec.~\ref{sec:exact_diagonalization}) transition between the two limits. Similarly, the light green markers and shading are the corresponding regions from a numerical perturbation series in $\Gamma$ (Sec.~\ref{subsec:numerical_treatment}). The transitions sharpen as $N$ increases, and a finite-size scaling analysis indicates that they become sharp in the thermodynamic limit. The energy density around which the transitions sharpen is consistent within numerical error amongst all order parameters. Thus we identify this energy density, once properly extrapolated to infinite $N$, as $\epsilon_c (\Gamma)$. It is difficult to make the extrapolation from accessible system sizes, and we observe a slight finite-size drift in the transition towards smaller $|\epsilon|$ as $N$ increases. For this reason, the marked points in Fig.~\ref{fig:micro_phase} are actually lower bounds for $\epsilon_c (\Gamma)$. 

Expanding the eigenstates perturbatively in $\Gamma$ sets upper bounds. A rough analytical treatment of the perturbation series sets a clear upper bound for $\epsilon_c (\Gamma)$ at $-\Gamma / e$ (Sec.~\ref{subsec:rough_estimates}). Proceeding more carefully but still analytically (Sec.~\ref{subsec:single_resonance_approximation}), we obtain the green dashed curve in Fig.~\ref{fig:micro_phase}. This curve is still only an upper bound on $\epsilon_c$, but it behaves as $-\Gamma$ at small $\Gamma$.

$\epsilon_c (\Gamma)$ lies between the green markers and the green dashed curve in Fig.~\ref{fig:micro_phase}. In addition, we conjecture that the bound $\epsilon_c \leq -\Gamma$ as $\Gamma \rightarrow 0$ is tight, and conjecture that
\begin{equation} \label{eq:phase_boundary_conjecture}
\epsilon_c (\Gamma) = - \Gamma 
\end{equation}
for finite $\Gamma$ as well. This is the black dashed line in Fig.~\ref{fig:micro_phase}. J\"{o}rg et al. observed~\cite{jorg2008simple} that the QREM ground state undergoes a quantum phase transition between the REM ground state and $\ket{\rightarrow \cdots \rightarrow}$ at $\Gamma = -\epsilon_0$ ($\epsilon_0$ given by Eq.~\eqref{eq:classical_T_c}). Our conjecture for $\epsilon_c (\Gamma)$ implies that the dynamical phase boundary is consistent with this result. It is consistent with all of our numerical and analytical bounds as well.

The critical eigenstates, i.e., those with $\epsilon = \epsilon_c (\Gamma)$, have order parameter values intermediate between the localized and ergodic limits. We are able to fit the dimensionless order parameters to a finite-size scaling ansatz $y = y_c + f(N^{\frac{1}{\tilde{\nu}}} (\epsilon - \epsilon_c))$ ($y$ is the order parameter). The critical exponent $\tilde{\nu} = 0.4(1)$, independent of $\Gamma$. The critical amplitudes $y_c$, though, do depend on $\Gamma$. This could be due to finite-size effects, but if not, the critical eigenstates are described by a line of fixed points.

Finally, it is interesting to compare the QREM's many-body mobility edge to that of a transition between ensembles of random matrix theory (RMT).  In particular, the random symmetric matrix $\hat{H} = \hat{A} + v \hat{G}$, where $\hat{A}$ is a diagonal random matrix whose elements are i.i.d. random variables of order 1 and $\hat{G}$ is a Gaussian random matrix, has been studied extensively as a model for the crossover from Poisson to GOE statistics. The extent to which $\hat{H}$ exhibits Poisson versus GOE statistics is determined by the ratio of $\hat{A}$'s mean level spacing, denoted $s$, to the typical off-diagonal matrix elements. If $\hat{A}$ has eigenvalues distributed uniformly over a fixed interval, $s \sim 1/D$ for matrix dimension $D$ and thus the scale on which $\hat{H}$'s level statistics transition from Poisson to GOE is $v \sim 1/D$.

We, however, are interested in when $\hat{A}$'s eigenvalues are distributed according to Eq.~\eqref{eq:distribution}. The mean level spacing $s$ is well-defined locally, but it depends on the local energy density $\epsilon$. In particular,
\begin{equation} \label{eq:level_spacing_scaling}
s(\epsilon) \sim \sqrt{N} e^{-N(\ln{2} - \epsilon ^2)}.
\end{equation}
By setting $v \sim s(\epsilon)$, we see a smooth interpolation from Poisson to GOE statistics at the energy density $\epsilon$. Yet for this choice of scaling for $v$, all other energy densities flow exponentially fast towards one of the limiting statistics as $N$ increases. The phase boundary is horizontal in the $v - \epsilon$ plane. This is in marked contrast to the QREM, where we find a non-trivial $\epsilon_c(\Gamma)$. The locality of the transverse field in Eq.~\eqref{eq:Hamiltonian} thus plays an important role in the physics of this transition.

\section{Exact Diagonalization} \label{sec:exact_diagonalization}

The most direct way to study the QREM is through exact diagonalization. We generate between $200$ and $20,000$ realizations of the Hamiltonian in Eq.~\eqref{eq:Hamiltonian} for system sizes ranging from $N=8$ to $14$ and $\Gamma$ ranging from $0.10$ to $0.40$. We obtain the entire spectrum and complete set of eigenstates through full exact diagonalization. The eigenstates are then binned according to energy density so as to study how average properties depend on energy.

We use three types of averages, often simultaneously. Quantum-mechanical averages within single eigenstates are denoted by angular brackets $\avg{\cdots}$, averages among eigenstates of a single sample within an energy density window are denoted by a bar $\overline{\cdots}$, and averages between realizations of disorder are denoted by square brackets $[\cdots]$. Unless noted otherwise, every quantity that we describe in this section is averaged both within an energy density window and over disorder.

\subsection{Spectral Statistics} \label{subsec:spectral_statistics}

The most straightforward numerical approach to the localization-delocalization transition is to study the statistical properties of the spectra. Each spectrum can be split into two regions: Wigner-Dyson statistics dominate the distribution of energy levels for $E$ close to $0$, whereas Poisson statistics dominate for $E$ far from $0$. From a random-matrix-theory perspective, Wigner-Dyson statistics govern the spectra of matrices in which every element is an independent random variable (strictly speaking, the Wigner-Dyson distribution gives the probability of level spacings in GOE-distributed matrices). Poisson statistics, on the other hand, describe the energy gaps when the eigenvalues themselves are independent and uniformly distributed. As is commonly done, we identify Wigner-Dyson statistics with delocalized eigenstates and Poisson statistics with localized eigenstates. We find that the transition between these two is sharp in the thermodynamic limit. Thus we identify the large-$|E|$ region as a localized phase and the small-$|E|$ region as a delocalized phase.

To be more quantitative, we use three measures of the energy gap distribution. First, we calculate the level spacing ratio~\cite{oganesyan2007localization}. We define
\begin{equation} \label{eq:r_definition}
r_n \equiv \min \{ {\frac{E_{n+1} - E_{n}}{E_{n} - E_{n-1}}, \frac{E_{n} - E_{n-1}}{E_{n+1} - E_{n}}} \}
\end{equation}
and consider the average, $[\overline{r}]$. $[\overline{r}]$ represents the degree of level repulsion in the system: there is less variation in the separation between energy levels, and thus a larger $[\overline{r}]$, when those levels repel each other. The Poisson distribution, which corresponds to independent energy levels, has $[\overline{r}] \approx 0.39$. Compare this to the Wigner-Dyson distribution, for which level repulsion is significant and $[\overline{r}] \approx 0.53$. The values of $[\overline{r}]$ that we obtain from (\ref{eq:Hamiltonian}) lie between these two limits and tell us the strength of level repulsion at the target energies. We in turn interpret this as the degree of delocalization.

In addition to $[\overline{r}]$, we determine the cumulative distribution function of level spacings $s$. We characterize the CDF by two quantities used in the literature~\cite{zharekeshev1995scaling}: $I_1 \equiv P(s < 0.473[\overline{s}])$, i.e, the fraction of spacings that are less than (roughly) half the mean, and $I_2 \equiv P(s > 2.002[\overline{s}])$, i.e, the fraction that are greater than twice the mean. Both these values are much smaller in the Wigner-Dyson distribution than in the Poisson distribution, since level repulsion suppresses the appearance both of packed and isolated eigenvalues. Thus $I_1$ and $I_2$, like $[\overline{r}]$, quantify the degree of level repulsion, $I_1$ by examining the frequency of small gaps and $I_2$ by examining the frequency of large gaps. We present each as a deviation from Wigner-Dyson statistics relative to that of the Poisson distribution, i.e., we report
\begin{equation} \label{eq:J_1_definition}
J_1 \equiv \frac{I_1 - I_1^{(\textrm{WD})}}{I_1^{(\textrm{P})} - I_1^{(\textrm{WD})}}
\end{equation}
and
\begin{equation} \label{eq:J_2_definition}
J_2 \equiv \frac{I_2 - I_2^{(\textrm{WD})}}{I_2^{(\textrm{P})} - I_2^{(\textrm{WD})}},
\end{equation}
where the superscript $(\textrm{WD})$ refers to Wigner-Dyson statistics and the superscript $(\textrm{P})$ refers to Poisson statistics.

\begin{figure}[htbp]
\begin{center}
\includegraphics[width=0.96\columnwidth]{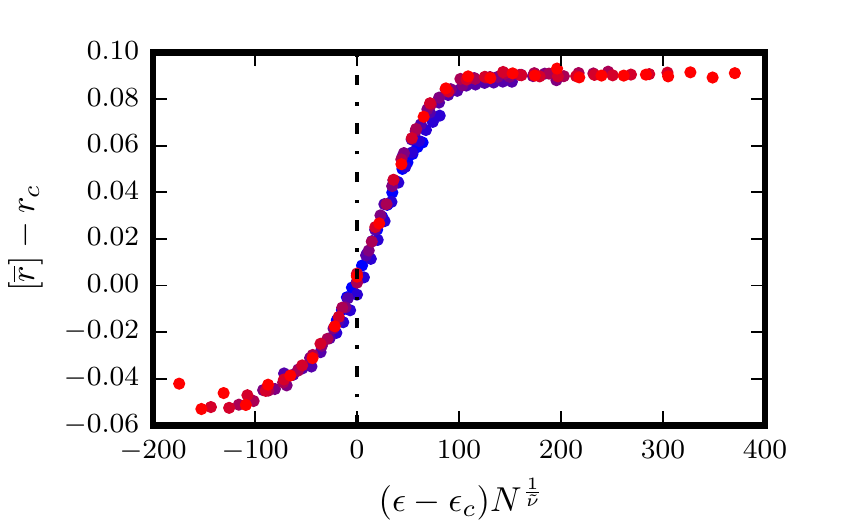}
\includegraphics[width=0.96\columnwidth]{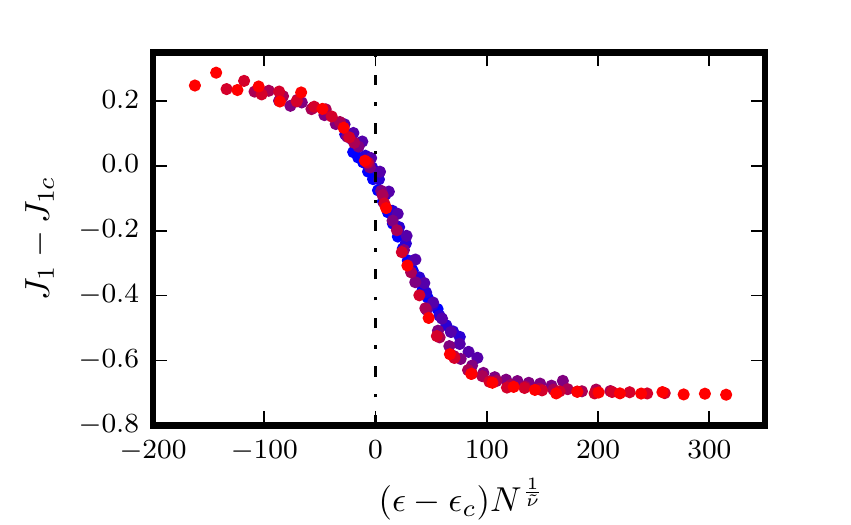}
\includegraphics[width=0.96\columnwidth]{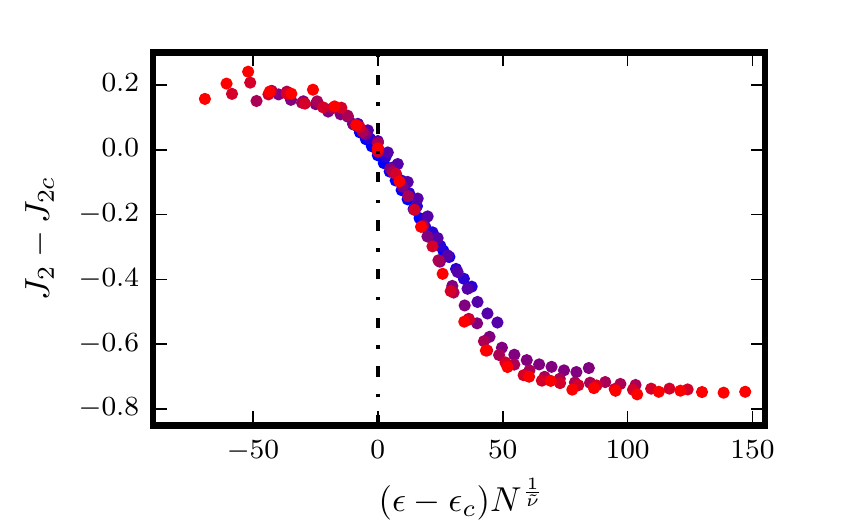}
\caption{Properties of the QREM (probabilistic) spectrum. All data is at $\Gamma = 0.20$, and $N = 8$  (blue) to $14$ (red). Eigenstates are binned within energy density windows of size $0.02$. These curves have been collapsed using the scaling form $y  = y_c + f(N^{\frac{1}{\tilde{\nu}}} (\epsilon - \epsilon_c))$. (a) The disorder-averaged level spacing ratio. $\epsilon_c = -0.34(2)$, $\tilde{\nu} = 0.39(9)$, and $r_c = 0.44(1)$. (b) The deviation of the frequency of smaller-than-average energy gaps from GOE statistics. $\epsilon_c = -0.33(2)$, $\tilde{\nu} = 0.41(9)$, and $J_{1c} = 0.73(1)$. (c) The deviation of the frequency of larger-than-average energy gaps from GOE statistics. $\epsilon_c = -0.34(2)$, $\tilde{\nu} = 0.46(9)$, and $J_{2c} = 0.80(1)$.}
\label{fig:ED_summary}
\end{center}
\end{figure}

\begin{figure}[htbp]
\begin{center}
\includegraphics[width=1.0\columnwidth]{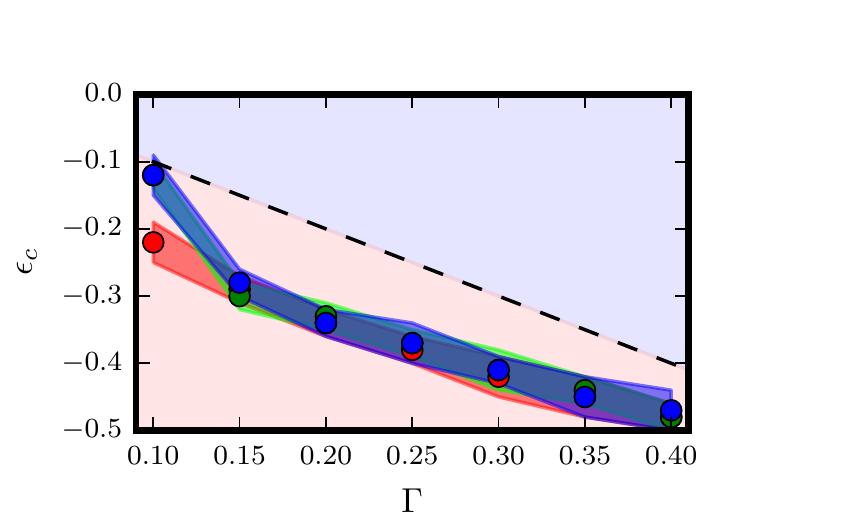}
\includegraphics[width=1.0\columnwidth]{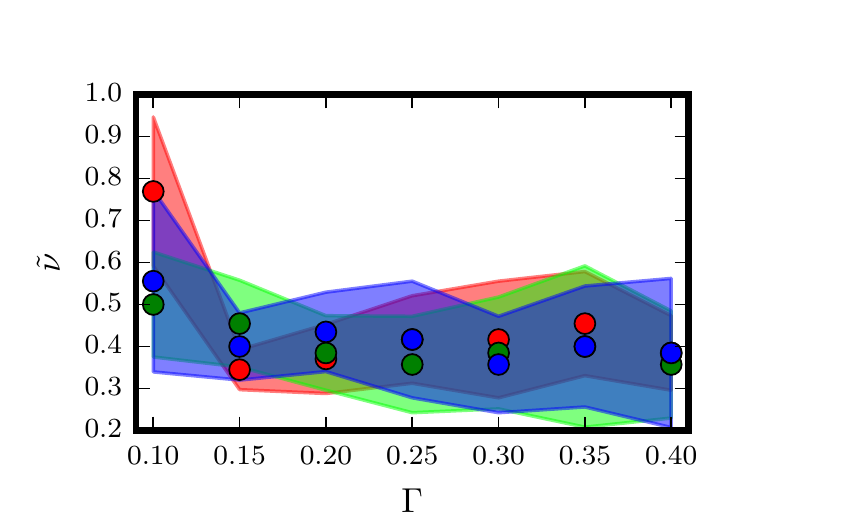}
\includegraphics[width=1.0\columnwidth]{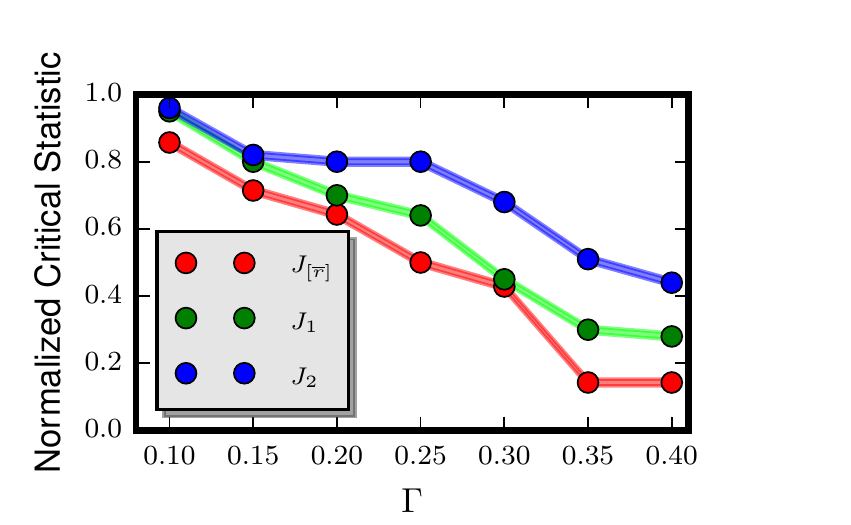}
\caption{Finite-size scaling parameters for the level spacing statistics as a function of $\Gamma$. The shaded bands indicate the range of parameter values over which the data collapses well. Red corresponds to $[\overline{r}]$, green corresponds to $J_1$, and blue corresponds to $J_2$. (a) The critical energy density, overlaid on a portion of the phase diagram of Fig.~\ref{fig:micro_phase}. 
(b) The critical exponent $\tilde{\nu}$. 
(c) The value of the level spacing statistic at criticality. We plot $([\overline{r}] - r^{(\textrm{WD})})/(r^{(\textrm{P})} - r^{(\textrm{WD})})$ and denote it by $J_{[\overline{r}]}$, in analogy with $J_1$ and $J_2$.} 
\label{fig:ED_scaling}
\end{center}
\end{figure}

Fig.~\ref{fig:ED_summary} shows $[\overline{r}]$, $J_1$, and $J_2$ at various $N$ as a function of energy density $\epsilon$ (at $\Gamma = 0.20$, a representative field strength), successfully collapsed using the scaling form $y  = y_c + f(N^{\frac{1}{\tilde{\nu}}} (\epsilon - \epsilon_c))$. From these we obtain $\epsilon_c$ and $\tilde{\nu}$, which are shown as functions of $\Gamma$ in Fig.~\ref{fig:ED_scaling}a,b. Except for at $\Gamma = 0.10$, all three statistics predict a common $\epsilon_c$ and $\Gamma$-independent $\tilde{\nu}$. Fig.~\ref{fig:ED_scaling}c shows the corresponding level statistic values at $\epsilon_c$, which we obtain from the collapsed curves. The statistics become more Wigner-Dyson-like as $\Gamma$ increases. It is in this sense that the character of the transition depends on $\Gamma$.

\subsection{Local Eigenstate Observables} \label{subsec:local_eigenstate_observables}

\begin{figure}[htbp]
\begin{center}
\includegraphics[width=1.0\columnwidth]{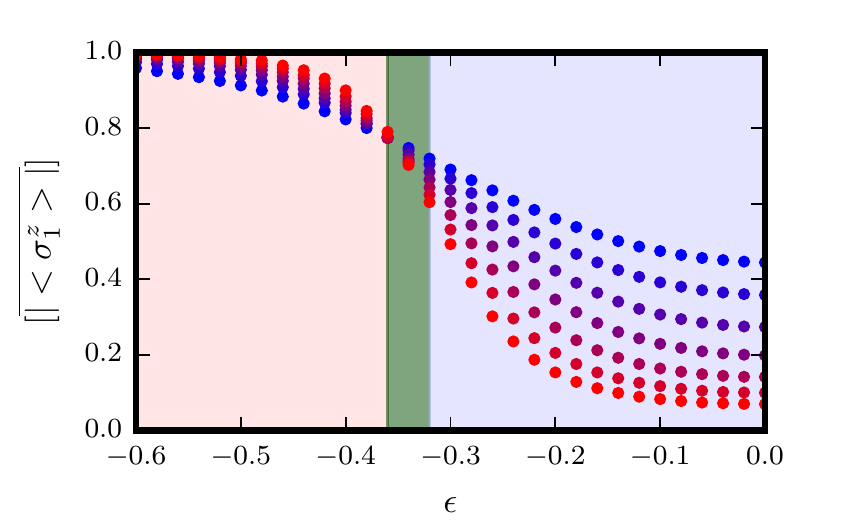}
\caption{The disorder-averaged magnitude of the eigenstate single-spin magnetization, as determined via exact diagonalization, as a function of the eigenstate energy density. These results are taken at $\Gamma = 0.20$, from $N=8$ (blue) to $N=14$ (red). Eigenstates are binned in energy density windows of size $0.02$. The background shading corresponds to the predicted phase at that energy density: red is localized, blue is delocalized, and green is the transition region as determined by the QREM's spectral statistics.}
\label{fig:ED_avg_mag}
\end{center}
\end{figure}

The magnetization of single spins in each phase clearly distinguishes the two. Since a delocalized eigenstate has $\avg{\hat{\sigma}_1^z} = 0$ and a localized eigenstate has $\avg{\hat{\sigma}_1^z} = \pm 1$, we consider $|\avg{\hat{\sigma}_1^z}|$. Fig~.\ref{fig:ED_avg_mag} shows the values as a function of $\epsilon$. We see clear finite-size flow towards $0$ in the delocalized phase and towards $1$ in the localized phase. The crossover region's location is consistent with the spectral statistics.

\begin{figure}[htbp]
\begin{center}
\includegraphics[width=1.0\columnwidth]{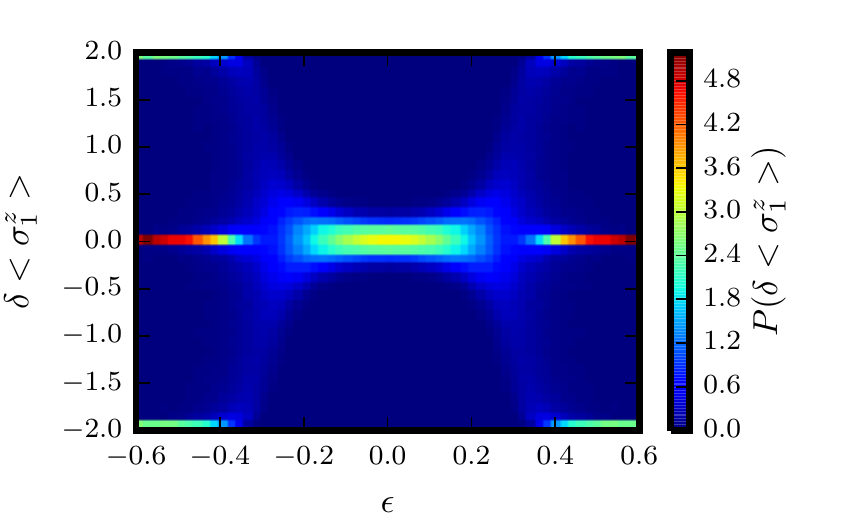}
\caption{Probability density function for the difference between single-spin magnetizations of spectrally-adjacent eigenstates (y-axis), over a range of energy densities (x-axis). Each vertical slice is a separate probability distribution. These distributions are at $\Gamma = 0.20$ and $N = 14$. We used energy density windows of $0.02$ for each distribution.}
\label{fig:ED_spider_diagram}
\end{center}
\end{figure}

The limit $[\overline{|\avg{\hat{\sigma}_1^z}|}] \rightarrow 1$ in the localized phase violates the ETH. We further demonstrate that the localized eigenstates fail to thermalize by studying how $\avg{\hat{\sigma}_1^z}$ fluctuates from state to state within a sample. More precisely, we define 
\begin{equation} \label{eq:delta_mag_definition}
\delta \avg{\hat{\sigma}_1^z}^{(n)} \equiv \braket{n+1 | \hat{\sigma}_1^z}{n+1} - \braket{n | \hat{\sigma}_1^z}{n}.
\end{equation}
Fig.~\ref{fig:ED_spider_diagram} shows the distribution of $\delta \avg{\hat{\sigma}_1^z}^{(n)}$, for all $n$ within an energy density window, as a function of $\epsilon$. In the localized phase, the distributions have weight at $\pm 2$. The total weight at $\pm 2$ is as much as at $0$, signifying that $\avg{\hat{\sigma}_1^z}$ is as likely to switch sign from one eigenstate to the next as it is to not. In the delocalized phase, the entire weight is centered around $0$. This is further evidence that each individual delocalized eigenstate is thermal.

\subsection{Connected Autocorrelations} \label{subsec:connected_autocorrelations}

The operators $\{\hat{\sigma_i^z}\}$ evolve over time as a result of the transverse field, so  correlations in time are important characteristics of the two phases. We quantify this by studying the connected autocorrelation function
\begin{equation} \label{eq:autocorrelation_definition}
\begin{split}
\avg{\hat{\sigma}_1^z(t) \hat{\sigma}_1^z(0)}_C^{(n)} \equiv &\braket{n | \hat{\sigma}_1^z(t) \hat{\sigma}_1^z(0)}{n} \\
&- \braket{n | \hat{\sigma}_1^z(t)}{n} \braket{n | \hat{\sigma}_1^z(0)}{n}.
\end{split}
\end{equation}
This quantity is computed exactly from the full exact diagonalization, up to $t_{\textrm{max}} = 90$ in steps of $0.5$. See Fig.~\ref{fig:ED_time_evolution} for the results. $[\overline{\avg{\hat{\sigma}_1^z(t) \hat{\sigma}_1^z(0)}_C}]$ is sufficiently close to 0 at all times in the localized phase that it is hard to extract a meaningful decay. This is consistent with the localized eigenstates being weakly dressed single configurations of spins with magnetization $\propto 1-\frac{1}{N^2}$. 

\begin{figure}[htbp]
\begin{center}
\includegraphics[width=1.0\columnwidth]{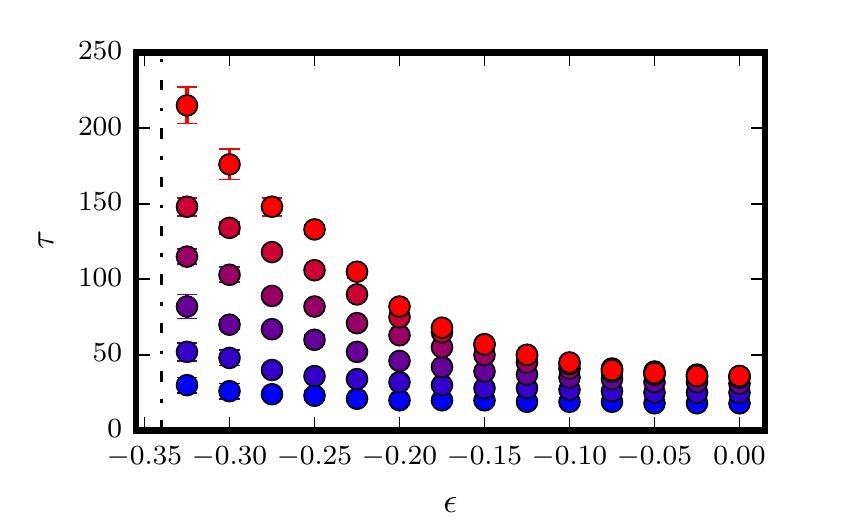}
\includegraphics[width=1.0\columnwidth]{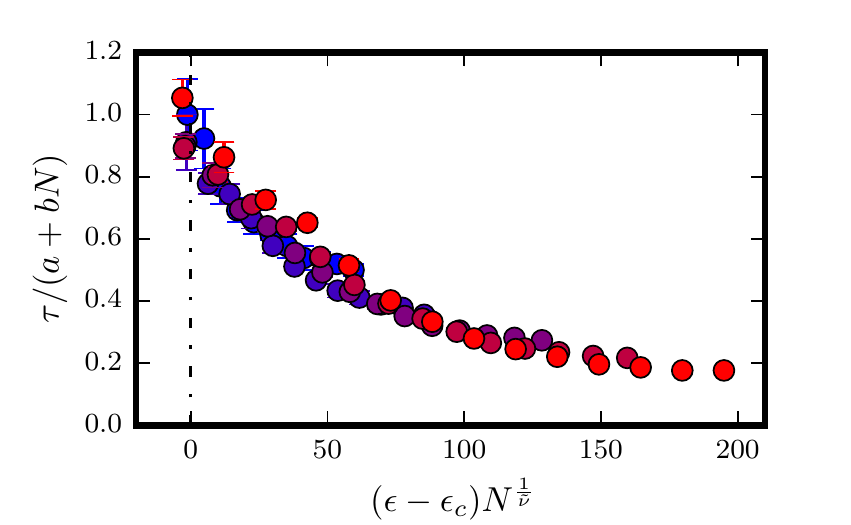}
\caption{The decay time of the eigenstate connected autocorrelation function $[\overline{\avg{\sigma_1^z(t) \sigma_1^z(0)}_C}]$, as a function of energy density. These results are taken at $\Gamma = 0.20$, $N = 8$ (blue) to $N = 13$ (red). The binning in energy density is $\Delta \epsilon = 0.025$. (a) The unscaled decay times. The vertical dashed line marks the critical energy density as determined by the QREM's spectral statistics. (b) Collapsed curves (omitting $N = 8$), using the form $\tau  = (a + bN) f(N^{\frac{1}{\tilde{\nu}}} (\epsilon - \epsilon_c))$. $\epsilon_c = -0.32(2)$, $\tilde{\nu} = 0.40(7)$, $a = -2.9(1) \cdot 10^{2}$, and $b = 38(2)$.}
\label{fig:ED_decay_times}
\end{center}
\end{figure}

In the delocalized phase, $[\overline{\avg{\hat{\sigma}_1^z(t) \hat{\sigma}_1^z(0)}_C}]$ is initially close to 1 and decays exponentially to 0. The decay becomes slower as $\epsilon$ nears the transition.
We fit the long-time behavior of $\ln{[\overline{\avg{\sigma_1^z(t) \sigma_1^z(0)}_C}]}$ to a straight line and extract the slope. From this we study how the decay time $\tau$ depends on $\epsilon$. See Fig.~\ref{fig:ED_decay_times}. At $\epsilon \approx 0$, the decay time saturates exponentially with $N$ to a finite value (not shown). As $\epsilon \rightarrow \epsilon_c$, $\tau$ increases monotonically. Very close to $\epsilon_c$, $\tau$ is diverging with system size. We obtain good scaling collapse with the form $\tau  = (a + bN) f(N^{\frac{1}{\tilde{\nu}}} (\epsilon - \epsilon_c))$. This form captures $\tau$'s linear dependence on $N$ near the transition. We again find that $\epsilon_c$ and $\tilde{\nu}$ agree with the spectral statistics.

\subsection{Eigenstate Structure} \label{subsec:eigenstate_structure}

\begin{figure}[t]
\begin{center}
\includegraphics[width=1.0\columnwidth]{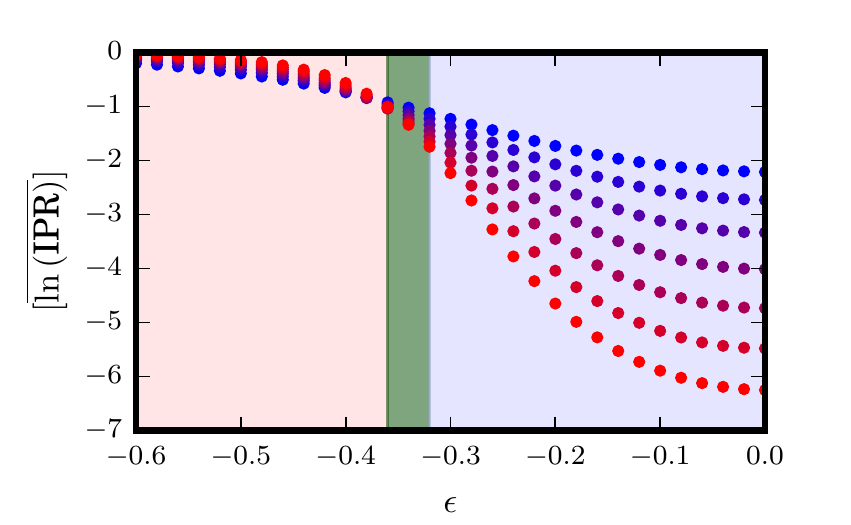}
\caption{The disorder-averaged order of magnitude of the eigenstate IPR from exact diagonalization, divided by system size. These results are taken at $\Gamma = 0.20$ for sizes $N=8$ (blue) to $N=14$ (red). Eigenstates are binned in energy density windows of size $\delta \epsilon = 0.02$. The background shading corresponds to the predicted phase at that energy density: red is localized, blue is delocalized, and green is the transition region as determined by the spectral statistics $[\overline{r}]$.}
\label{fig:ED_avg_ipr}
\end{center}
\end{figure}

As mentioned in Section~\ref{subsec:dynamical_phase_diagram}, configurations of $N$ spin-1/2s map to the corners of an $N$-dimensional hypercube. The $\{ \hat{\sigma}_i^z \}$ basis is then the ``coordinate'' basis, and we consider diagnostics from Anderson localization \cite{evers08}. These have already been used in the MBL context \cite{buccheri2011structure,de2013ergodicity}, particularly the inverse participation ratio (IPR), defined as
\begin{equation} \label{eq:IPR_definition}
Y_2 \equiv \sum _{\{\sigma_i^z\}} |\braket{\{ \sigma_i^z \}}{\psi}|^4.
\end{equation}
The IPR quantifies how many sites of the hypercube $\ket{\psi}$ is distributed over. $Y_2 = 1$ corresponds to $\ket{\psi}$ concentrated on a single site and $Y_2 = 2^{-N}$ corresponds to $\ket{\psi}$ with equal weight on all sites. Fig.~\ref{fig:ED_avg_ipr} shows $[\overline{\ln{Y_2}}]$, a measure of $Y_2$'s typical order of magnitude. $[\overline{\ln{Y_2}}]$ flows towards 0 with system size in the localized phase, signifying that the eigenstates are concentrated onto single sites. $[\overline{\ln{Y_2}}]$ decreases linearly with $N$ in the delocalized phase, signifying that the eigenstates extend over Hamming distances of order $N$. As with $[\overline{|\avg{\hat{\sigma}_1^z}|}]$, there is a well-defined crossover region consistent with the spectral statistics.

\begin{figure}[t]
\begin{center}
\includegraphics[width=1.0\columnwidth]{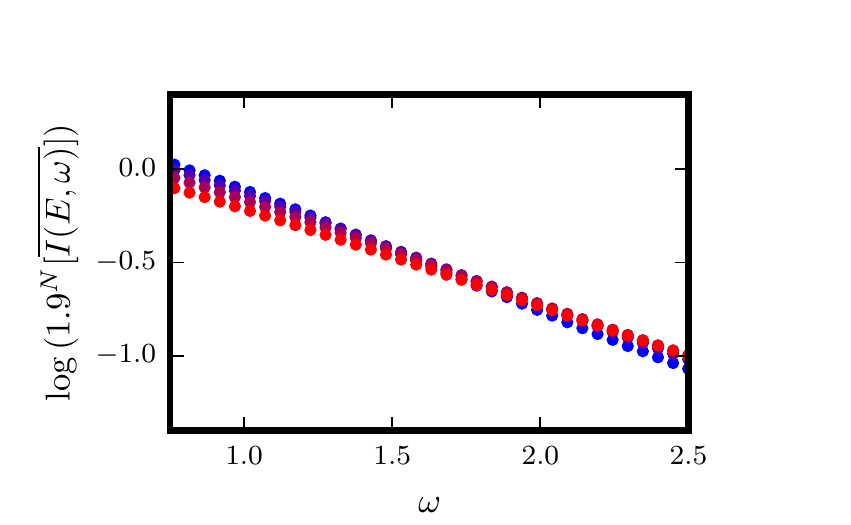}
\caption{Semi-log plot of probability overlaps between eigenstates separated by energy $\omega$, in the delocalized phase ($\epsilon = -0.10$, $\Gamma = 0.40$) for $N=8$ (blue) to $N=14$ (red). This data has been averaged over states within an energy window $\delta \epsilon = 0.05$ centered on $\epsilon = -0.10$, and averaged over samples. The lines have been shifted vertically to illustrate their dependence on $N$. Although the slopes of the lines depend slightly on $N$, the rough value is $-0.50$.}
\label{fig:ED_del_prob_overlap}
\end{center}
\end{figure}

We also compute the eigenstate correlation function $I(E, \omega)$, defined as
\begin{equation} \label{eq:ECF_definition}
I(E, \omega) = \sum_{m, n} \delta_{E, E_m} \delta_{E + \omega, E_n} \sum_{\{\sigma_i^z\}} |\psi_m(\{\sigma_i^z\})|^2 |\psi_n(\{\sigma_i^z\})|^2
\end{equation}
where the sum over $m$ and $n$ is over eigenstates. This quantity measures the overlap between eigenstates' probabilities as a function of their location in the spectrum and energy separation. $I(E, \omega)$ is known as the two-particle spectral function in the context of finite-dimensional lattices \cite{chalker1988scaling}. There, $I(E, \omega)$ decays exponentially with $\omega$ for $E$ and $E + \omega$ both in the delocalized phase. The scale on which $I(E, \omega)$ decays is the Thouless energy $\omega_0 = D/L^2$ ($D$ is the diffusion coefficient and $L$ is the linear system size). Fig.~\ref{fig:ED_del_prob_overlap} shows the same behavior in the QREM's delocalized phase. The characteristic scale is independent of $N$, which is reasonable since our fully-connected model has an effective ``linear size'' of 1. $[\overline{I(E, \omega)}]$ also decays exponentially with $N$, for the same reason that $Y_2$ does. In the localized phase, $I(E, \omega)$ decays as $2^{-N} \omega^{-2}$. We justify this in Section~\ref{sec:naive_perturbation_theory}, where we apply first-order perturbation theory to this model. The results in Fig.~\ref{fig:ED_loc_prob_overlap} agree remarkably well with Eq.~\eqref{eq:avg_first_order_overlap} below.

\begin{figure}[t]
\begin{center}
\includegraphics[width=1.0\columnwidth]{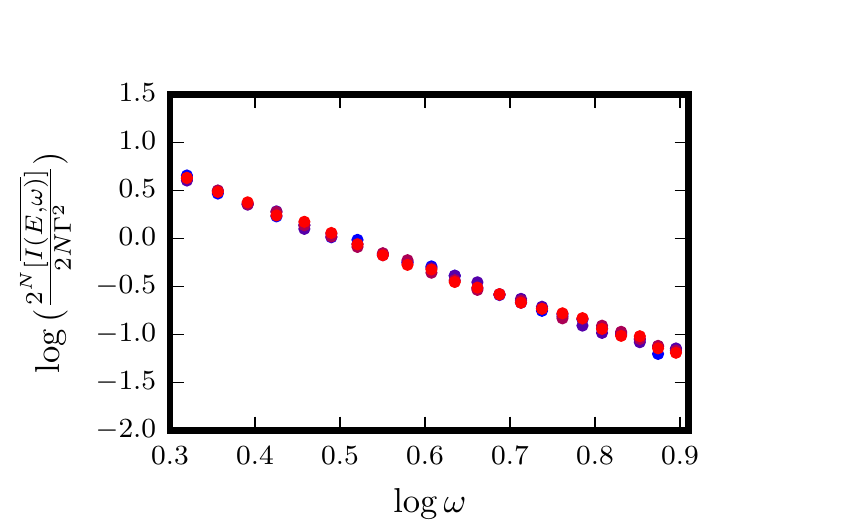}
\caption{Log-log plot of probability overlaps between eigenstates separated by  energy $\omega$, in the localized phase ($\epsilon = -0.20$, $\Gamma = 0.10$) for $N=8$ (blue) to $N=14$ (red).  This data has been averaged over states within an energy window $\delta \epsilon = 0.05$ centered on $\epsilon = -0.20$, and averaged over samples. The lines have been shifted vertically to illustrate their dependence on $N$. The lines all have slopes of $-2$, and are well-described by perturbation theory (see Eq.~\eqref{eq:avg_first_order_overlap}).}
\label{fig:ED_loc_prob_overlap}
\end{center}
\end{figure}

\section{Naive Perturbation Theory} 
\label{sec:naive_perturbation_theory}

The QREM eigenstates differ drastically across a well-defined boundary. The spectral statistics and eigenstate properties characterize and distinguish the two phases. However, the numerical results by themselves are somewhat opaque and limited to finite sizes. 
We get a more transparent picture of the MBLD transition by considering the perturbative structure. 
Section~\ref{sec:forward_scattering_analysis} systematically investigates the perturbative-in-$\Gamma$ expansion of the eigenstates. 
Here we study the first-order corrections heuristically and compare to the exact diagonalization results for the localized phase.

To first order in $\Gamma$, each QREM eigenstate has probability amplitude 1 on its initial site of the hypercube (with unperturbed energy $E_0$) and probability amplitude $-\frac{\Gamma}{E_0 - E_j}$ on the $j$'th adjacent corner. All other sites have probability amplitude 0. For the IPR and $\avg{\sigma _1^z}$ of such an eigenstate, we obtain:
\begin{equation} \label{eq:first_order_IPR_result}
\textrm{IPR} = \frac{1 + \sum _{j=1}^N \frac{\Gamma^4}{(E_0 - E_j)^4}}{\left( 1 + \sum _{j=1}^N \frac{\Gamma^2}{(E_0 - E_j)^2} \right) ^2},
\end{equation}
\begin{equation} \label{eq:first_order_mag_result}
\avg{\sigma_1^z} = 1 - \frac{2 \Gamma ^2}{(E_0 - E_1)^2}.
\end{equation}

We now study the distributions of Eq.~\eqref{eq:first_order_IPR_result} and Eq.~\eqref{eq:first_order_mag_result} over samples, treating $E_0$ as a fixed parameter of order $N$. With probability 1 (in the thermodynamic limit), the $N$ random variables $E_j$ are all $O(\sqrt{N})$ and we expand $\frac{1}{E_0 - E_j}$ in powers of $\frac{E_j}{E_0}$. This is not strictly allowed because, although its typical values are small, the moments of $-\frac{\Gamma}{E_0 - E_j}$ diverge. For now, we assume that the typical values of Eq.~\eqref{eq:first_order_IPR_result} and Eq.~\eqref{eq:first_order_mag_result} will be satisfactory estimates and so assume that all $E_j \ll E_0$.

To determine the disorder averages, we need only keep the first term that is even in all $E_j$. To determine the disorder variances, we in addition require the first term odd in $E_j$. Carrying this out,
\begin{equation} \label{eq:first_order_IPR_approx}
\textrm{IPR} \approx 1 - \frac{2 \Gamma^2}{N \epsilon_0^2} - \frac{4 \Gamma^2}{N^2 \epsilon_0^3} \sum _{j=1}^N \epsilon_j,
\end{equation}
\begin{equation} \label{eq:first_order_mag_approx}
\avg{\sigma_1^z} \approx 1 - \frac{2 \Gamma^2}{N^2 \epsilon_0^2} - \frac{4 \Gamma^2}{N^2 \epsilon_0^3} \epsilon_1,
\end{equation}
where $\epsilon_0 \equiv E_0/N$, $\epsilon_j \equiv E_j/N$. All $E_j$ have variance $N/2$, so all $\epsilon_j$ have variance $1/(2N)$ and $\sum _{j=1}^N \epsilon_j$ has variance $1/2$. Thus
\begin{equation} \label{eq:first_order_IPR_moments}
\begin{split}
&[\textrm{IPR}] = 1 - \frac{2 \Gamma^2}{N \epsilon_0^2}, \\
&\textrm{Var}(\textrm{IPR}) = \frac{8 \Gamma^4}{N^4 \epsilon_0^6},
\end{split}
\end{equation}
\begin{equation} \label{eq:first_order_mag_moments}
\begin{split}
&[\avg{\sigma_1^z}] = 1 - \frac{2 \Gamma^2}{N^2 \epsilon_0^2}, \\
&\textrm{Var}(\avg{\sigma_1^z}) = \frac{8 \Gamma^4}{N^5 \epsilon_0^6}.
\end{split}
\end{equation}
In addition, note that $[\ln{(\textrm{IPR})}] = [\textrm{IPR}] - 1$ at this order.

We also estimate $I(E_0, \omega)$ in this manner. Suppose that two eigenstates have energies $E_0$ and $E_0 + \omega$. In order for their wavefunctions to overlap at first order in $\Gamma$, their zeroth-order eigenstates must be on adjacent corners of the hypercube. This occurs in a fraction $N/2^N$ of the samples. When the eigenstates do overlap at first order,
\begin{equation} \label{eq:first_order_prob_overlap}
I(E_0, \omega) = 2 \frac{\Gamma^2}{\omega^2}.
\end{equation}
Thus
\begin{equation} \label{eq:avg_first_order_overlap}
[I(E, \omega)] = \frac{2N \Gamma^2}{2^N \omega^2}.
\end{equation}

These are simple estimates, and we can generalize them to higher orders. 
However, most turn out to be incompatible with our exact diagonalization results, even in how they scale with $N$ and $\epsilon_0$. 
The one exception is $[I(E_0, \omega)]$, for which Eq.~\eqref{eq:avg_first_order_overlap} is very accurate in the well-localized region. We expect that the overall discrepancy is due to our limited range of system sizes. In particular, we have expanded in $\frac{E_j}{E_0}$, which is $O(N^{-1/2})$ in typical samples. 
Yet even for our largest system ($N = 14$), $N^{-1/2} > 1/4$. Similarly, the exponentially small number of ``atypical'' samples is not that small. Eq.~\eqref{eq:avg_first_order_overlap} successfully describes $[I(E_0, \omega)]$ because there is no need to control the random variables $E_j$ at first order, since by definition $E_0 - E_j$ must be equal to $\omega$. When we remove complications arising from the randomness in $E_j$, the leading-order estimates become highly accurate, but sadly we cannot do this for any but $[I(E_0, \omega)]$.

With our limited systems sizes we must turn to more sophisticated approximation schemes to understand the localization-delocalization transition. We have two goals in doing so: to quantitatively predict the critical energy density $\epsilon_c$ in a procedure more efficient than exact diagonalization, and to understand how the structure of localized eigenstates changes as $\epsilon$ approaches $\epsilon_c$. The Forward-Scattering Approximation accomplishes both.

\section{Forward Scattering Analysis}
\label{sec:forward_scattering_analysis}

The Forward-Scattering Approximation (FSA) treats the transverse field perturbatively and keeps only the leading-order contribution to the wavefunction amplitude at each site (see \cite{huse2015localized,pietracaprina2015forward} for recent applications of the method to the Anderson problem). We approach this approximation from the hypercube perspective. The $E(\{\hat{\sigma}_i^z\})$ term in the Hamiltonian is a random on-site potential and the $\Gamma$ term allows hopping between adjacent sites. Without loss of generality, our unperturbed state has all $\sigma_i^z = -1$. Denote it by $\ket{- \dotsc -}$, with unperturbed energy $E_0$. The perturbed state is denoted $\ket{\psi}$ with energy $E_0 + \Delta$. From time-independent perturbation theory,
\begin{equation} \label{eq:perturbation_formalism}
\ket{\psi} = \ket{- \dotsc -} + \frac{P}{E_0 - E(\{\hat{\sigma}_i^z\})} \left(-\Gamma \sum_{i=1}^N \hat{\sigma}_i^x - \Delta \right) \ket{\psi}
\end{equation}
where $P \equiv I - \ket{- \dotsc -} \bra{- \dotsc -}$. The leading-order contributions to $\braket{\{\hat{\sigma}_i^z\}}{\psi}$ are the ``directed'' paths from $\ket{- \dotsc -}$ to $\ket{\{\hat{\sigma}_i^z\}}$, i.e.,
\begin{equation} \label{eq:FSA_expression}
\braket{\{\hat{\sigma}_i^z\}}{\psi} \equiv \psi (\{\hat{\sigma}_i^z\}) = \sum_p \prod_{i=1}^{l(\{\hat{\sigma}_i^z\})} \frac{\Gamma}{E_0 - E(p_i)},
\end{equation}
where the sum is over the $l(\{\hat{\sigma}_i^z\})!$ directed paths from the initial site to the site $\{\hat{\sigma}_i^z\}$ and $E(p_i)$ is the unperturbed energy of the $i$th site along a given path $p$.
We treat $E_0$ as a tunable parameter. All $E(p_i)$ are random variables distributed according to Eq.~\eqref{eq:distribution}.

Eq.~\eqref{eq:FSA_expression} is the forward-scattering approximation to the eigenstates of the REM in a transverse field. We use it to investigate signatures of delocalization. As long as the perturbation theory is valid, i.e., all $\psi (\{\hat{\sigma}_i^z\}) \ll 1$, localization persists. Strictly speaking, the appearance of order-1 $\psi (\{\hat{\sigma}_i^z\})$, which we refer to as ``resonances'', does not mean that the exact eigenstates are delocalized, only that Eq.~\eqref{eq:FSA_expression} cannot describe them. Regardless, we take the energy density at which resonances proliferate in the perturbed wavefunction as a strong lower bound for $\epsilon_c$.

Although the FSA is an approximation, as it stands, Eq.~\eqref{eq:FSA_expression} must still be evaluated numerically. We do so below, but first consider further approximations. In Subsection~\ref{subsec:rough_estimates} we make quick estimates as to where the perturbation series breaks down. We refine these estimates in Subsection~\ref{subsec:single_resonance_approximation}. In Subsection~\ref{subsec:numerical_treatment} we evaluate Eq.~\eqref{eq:FSA_expression} numerically. Finally, in Sec.~\ref{sec:replica_treatment} we briefly consider the replica treatment of the FSA, which reproduces many of the results obtained more directly in the previous sections.

\subsection{Rough Estimates} 
\label{subsec:rough_estimates}

From Eq.~\eqref{eq:distribution}, most configurations have energies of order $\sqrt{N}$. Thus most terms in Eq.~\eqref{eq:FSA_expression} are $\frac{\Gamma}{E_0}$ to leading order in $N$, and a rough estimate comes from replacing all terms with $\frac{\Gamma}{E_0}$. Then
\begin{equation} \label{eq:FSA_simplest_wavefunction}
\psi (\{\hat{\sigma}_i^z\}) = l(\{\hat{\sigma}_i^z\})! \left( \frac{\Gamma}{E_0} \right) ^{l(\{\hat{\sigma}_i^z\})} \approx \left( \frac{l(\{\hat{\sigma}_i^z\}) \Gamma}{e E_0} \right) ^{l(\{\hat{\sigma}_i^z\})}.
\end{equation}
As a function of $l$, $\psi (\{\hat{\sigma}_i^z\})$ first exceeds $1$ at $l = N$, if at all. We estimate $\epsilon_c$ by setting $\frac{N \Gamma}{e E_0} = 1$:
\begin{equation} \label{eq:FSA_simplest_transition}
\epsilon_c \le - \frac{\Gamma}{e}.
\end{equation}

We have written the estimate above as a bound on the position of the transition curve because it clearly underestimates the possibility of small-denominator resonances. Even though most REM energies are of order $\sqrt{N}$, an exponentially-large number of states are still within any finite window near $E_0$. 
Such states contribute much more than $\frac{\Gamma}{E_0}$ to Eq.~\eqref{eq:FSA_expression}.  Nevertheless, this estimate agrees with ED in predicting that the transition lies at finite energy density/temperature. It also suggests that $\epsilon_c \rightarrow 0$ as $\Gamma \rightarrow 0$. 
Localization persists up to higher temperatures as the transverse field weakens.

This analysis is clearly invalid at $\epsilon_0 = 0$, where the weights $\frac{\Gamma}{-E(p_i)}$ are large for most sites. In this regime, the probability distributions for path contributions (the terms $\prod_{i=1}^{l(\{\hat{\sigma}_i^z\})} \frac{\Gamma}{E_0 - E(p_i)}$ in (\ref{eq:FSA_expression})) are long-tailed. Thus, we expect the largest-weighted path to dominate the sum over them. The ``greedy algorithm'' estimates this largest-weighted path as follows: most sites have REM energies of order $\sqrt{N}$, and the smallest energy among $m$ sites will be $O(\frac{\sqrt{N}}{m})$. 
The smallest energy of sites neighboring the initial site is $O(\frac{\sqrt{N}}{N})$, as the initial site has $N$ neighbors. The corresponding site has a neighbor at distance 2 with energy $O(\frac{\sqrt{N}}{N-1})$, for whom the smallest energy of neighboring sites at distance 3 is $O(\frac{\sqrt{N}}{N-2})$, etc., such that the largest path is typically
\begin{equation} \label{eq:FSA_high_T_wavefunction_v1}
\psi (\{\hat{\sigma}_i^z\}) \sim \frac{N\Gamma}{\sqrt{N}} \frac{(N-1)\Gamma}{\sqrt{N}} \cdots \frac{(N-l+1)\Gamma}{\sqrt{N}}.
\end{equation}
At large $N$,
\begin{equation} \label{eq:FSA_high_T_wavefunction_v2}
\psi (\{\hat{\sigma}_i^z\}) \sim (\Gamma \sqrt{N})^l.
\end{equation}
The wavefunction amplitude decays exponentially with distance for $\Gamma < \Gamma_c$, where
\begin{equation} \label{eq:FSA_high_T_crossover}
\Gamma_c = \frac{1}{\sqrt{N}}.
\end{equation}
Any non-zero $N$-independent field causes the perturbation series to break down. The crossover region goes as $N^{-1/2}$, and once $\Gamma$ exceeds $\Gamma_c$, the breakdown occurs immediately, i.e., at $l = O(1)$.

\subsection{Single Resonance Approximation} \label{subsec:single_resonance_approximation}

Consider the probability that a site at distance $l$ produces $\psi (\{\hat{\sigma}_i^z\}) > 1$, given that the closer sites all have energies of order $\sqrt{N}$. This checks the consistency of the treatment above, and we expect single sites to produce resonances in this way because the distribution of $\frac{\Gamma}{E_0 - E(p_i)}$ is long-tailed. From Eq.~\eqref{eq:FSA_simplest_wavefunction}, all sites at a distance $l-1$ have amplitude
\begin{equation}
\psi_{l-1} = (l-1)! \left( \frac{\Gamma}{E_0} \right) ^{l-1}.
\end{equation}
Then the site $\{\hat{\sigma}_i^z\}$ at distance $l$ has amplitude
\begin{equation}
\begin{split}
\psi (\{\hat{\sigma}_i^z\}) &= l \psi_{l-1} \left( \frac{\Gamma}{E_0 - E(\{\hat{\sigma}_i^z\})} \right) \\
&= l! \left( \frac{\Gamma}{E_0} \right) ^{l-1} \left( \frac{\Gamma}{E_0 - E(\{\hat{\sigma}_i^z\})} \right).
\end{split}
\end{equation}
The probability of $|\psi (\{\hat{\sigma}_i^z\})| > 1$ is
\begin{equation} \label{eq:single_probability}
\begin{split}
p_l &= \int _{E_0 \left( 1 - l!\frac{\Gamma^l}{E_0^l} \right)} ^{E_0 \left( 1 + l!\frac{\Gamma^l}{E_0^l} \right)} \mathrm{d}E \frac{1}{\sqrt{\pi N}} e^{-\frac{E^2}{N}} \\
&\approx 2 \epsilon \left( \frac{l\Gamma}{eE_0} \right) ^l \sqrt{\frac{N}{\pi}} e^{-N\epsilon^2}.
\end{split}
\end{equation}
Since the $E(\{\hat{\sigma}_i^z\})$ are independent, the $\psi(\{\hat{\sigma}_i^z\})$ for all sites at distance $l$ are (under the current assumptions) independent, and thus the probability of \textit{no} resonances at distance $l$ is
\begin{equation} \label{eq:SRA_probability_result}
\begin{split}
&(1 - p_l)^{\binom{N}{l}} \approx e^{-{\binom{N}{l}} p_l}\\ 
&\approx \exp\left(-2\epsilon \left( \frac{x\Gamma}{e\epsilon} \right) ^{Nx} \sqrt{\frac{N}{\pi}} e^{-N\epsilon^2} \left( \frac{1}{x^{x}(1-x)^{1-x}} \right) ^N\right),
\end{split}
\end{equation}
where $x \equiv l/N$. In the large-$N$ limit, this is $\exp{(-ke^{Nf(x, \epsilon)})}$ with
\begin{equation} \label{eq:exponent_expression}
f(x, \epsilon) \equiv x \ln{\frac{x\Gamma}{e\epsilon}} - \epsilon ^2 - x\ln{x} - (1-x)\ln{(1-x)}
\end{equation}
and $k = O(\sqrt{N})$. There will not be resonances in the thermodynamic limit and localization will persist if $f(x, \epsilon) < 0$ for all $x \in [0, 1]$. $f(x, \epsilon)$ is maximized at $x_{\textrm{max}} \equiv 1 - \epsilon / \Gamma$ with a value of
\begin{equation} \label{eq:IRA_equation}
f(x_{\textrm{max}}, \epsilon) = \ln{\frac{\Gamma}{e\epsilon}} + \frac{\epsilon}{\Gamma} - \epsilon^2.
\end{equation}
The zero of Eq.~\eqref{eq:IRA_equation} is at $\epsilon_c$. We find that
\begin{equation} \label{eq:IRA_approx_solution}
\epsilon_c = - \left(\Gamma - \sqrt{2} \Gamma ^2 + O(\Gamma^3)\right).
\end{equation}
The phase diagram in Fig.~\ref{fig:micro_phase} displays this estimate as a function of $\Gamma$. We get a much larger estimate than that in Eq.~\eqref{eq:FSA_simplest_transition}, yet $\epsilon_c$ is still proportional to $\Gamma$ as $\Gamma \rightarrow 0$.

In addition to an estimate for the critical energy density, this shows how suddenly and drastically the perturbation theory breaks down. Suppose we had defined a resonant site to be one whose amplitude exceeds $c$. Then the probability of no resonances at distance $l$ would be $\exp{(-\frac{k}{c}e^{Nf(x, \epsilon)})}$. The large-$N$ asymptotics only change if $c$ grows/decays faster than $ e^{\alpha N}$. Thus the resonances that appear at $\epsilon > \epsilon_c$ are exponentially large in system size, and all sites have exponentially small amplitudes at $\epsilon < \epsilon_c$.

We similarly study the infinite-temperature case, where the greedy algorithm estimated $\Gamma_c = N^{-1/2}$. As described above, we cannot consider single sites being resonant but we can consider single \emph{paths} being resonant. A path $p$ to a site at distance $l$ has amplitude
\begin{equation} \label{eq:single_path_amplitude}
|A| = \prod _{i=1} ^{l} \frac{\Gamma}{|E(p_i)|}.
\end{equation}
To simplify notation, we write this as
\begin{equation} \label{eq:path_clean_amplitude}
\ln{|A|} = l \ln{\frac{\Gamma}{\sigma}} + Y
\end{equation}
where
\begin{equation} \label{eq:path_notation_1}
\sigma \equiv \frac{\sqrt{\pi N}}{2},
\end{equation}
\begin{equation} \label{eq:path_notation_2}
Y \equiv \sum _{i=1} ^{l} \ln{\frac{\sigma}{|E(p_i)|}}.
\end{equation}

For $Y \gg 1$, $Y$ is distributed as
\begin{equation} \label{eq:path_distribution}
P(Y) = \frac{Y^{l-1}}{(l-1)!} e^{-Y}.
\end{equation}
Thus $|A| > 1$, i.e., $Y > Y_c \equiv l \ln{\frac{\sigma}{\Gamma}}$, occurs with probability
\begin{equation} \label{eq:single_path_resonance}
p_l = \int _{Y_c} ^{\infty} dY  \frac{Y^{l-1}}{(l-1)!} e^{-Y} = \frac{Y_c^{l-1}}{(l-1)!} e^{-Y_c} \left( 1 + O(Y_c^{-1}) \right).
\end{equation}
We use this estimate to determine the probability that none of the $\binom{N}{l}$ paths to sites at distance $l$ give resonances. The paths are not independent, but we assume that we can treat them independently in estimating whether $|A| \gg 1$ or $|A| \ll 1$. Then the probability of no resonances at distance $l$ is
\begin{equation} \label{eq:total_resonance_probability}
(1 - p_l)^{\binom{N}{l}} \approx e^{-{\binom{N}{l}} p_l}.
\end{equation}
Consider the exponent ${\binom{N}{l}} p_l \equiv c_l$, and in particular the ratio
\begin{equation} \label{eq:exponent_ratio}
\frac{c_{l+1}}{c_l} = \frac{N-l}{l+1} \left( 1 + \frac{1}{l} \right) ^l \frac{\Gamma}{\sigma} \ln{\frac{\sigma}{\Gamma}}.
\end{equation}

The right-hand side of Eq.~\eqref{eq:exponent_ratio} is decreasing as a function of $l$. When it drops below 1, the exponents in Eq.~\eqref{eq:total_resonance_probability}) decrease as $l$ increases, i.e., resonances become less likely as we move farther into the hypercube. In order to declare that resonances are unlikely throughout the entire hypercube, we require that $\frac{c_2}{c_1} < 1$. This is only possible when $\Gamma$ is such that
\begin{equation} \label{eq:Gamma_condition}
(N-1) \frac{\Gamma}{\sigma} \ln{\frac{\sigma}{\Gamma}} < 1.
\end{equation}
The critical $\Gamma$ is, to leading order,
\begin{equation} \label{eq:high_T_critical_Gamma}
\Gamma _c = \frac{\sigma}{N \ln{N}} = \frac{\sqrt{\pi}}{2 \sqrt{N} \ln{N}}.
\end{equation}
$\Gamma _c$ decays to 0 faster than the greedy estimate in Eq.~\eqref{eq:FSA_high_T_crossover} by a factor of $\ln{N}$, but otherwise Eq.~\eqref{eq:high_T_critical_Gamma} confirms the qualitative description. Perturbation theory inevitably breaks down in the thermodynamic, high-temperature limit at any non-zero $\Gamma$. We see the resonances immediately, i.e., at an $O(1)$ order in the expansion. This contrasts with the finite-temperature case, where the perturbative description persists until finite $\Gamma$ and the resonances appear after $O(N)$ terms in the series.

We have a description of when and how QREM eigenstates become non-perturbative in $\Gamma$, but we have made some uncontrolled approximations. Order-$1$ wavefunction amplitudes need not be due solely to single resonant sites or paths. This then implies that the amplitudes are not independent of each other. Below, we study the statistics of the wavefunctions in Eq.~\eqref{eq:FSA_expression} numerically, without any additional approximations. We check not only the estimate for the critical energy density but also whether the nature of the eigenstates agrees with the description above.

\subsection{Numerical Treatment} \label{subsec:numerical_treatment}

\begin{figure}[htbp]
\begin{center}
\includegraphics[width=1.0\columnwidth]{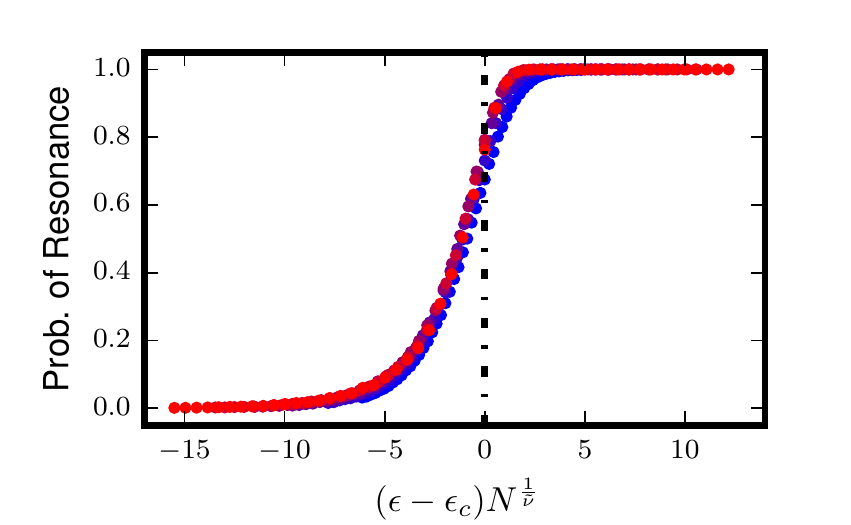}
\includegraphics[width=1.0\columnwidth]{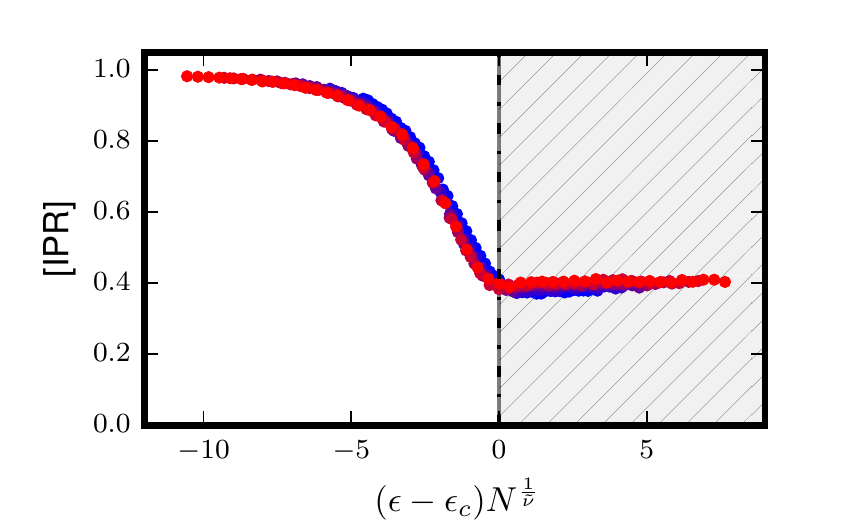}
\includegraphics[width=1.0\columnwidth]{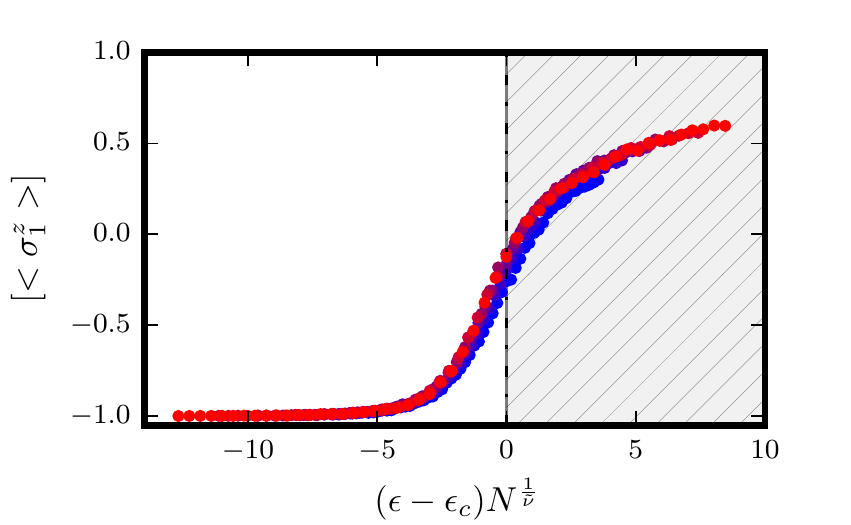}
\caption{Eigenstate properties as determined within the forward-scattering approximation, as a function of energy density. All data is at $\Gamma = 0.20$, and $N = 10$ (blue) to $20$ (red). These curves have been collapsed using the scaling form $y = f(N^{\frac{1}{\tilde{\nu}}} (\epsilon - \epsilon_c))$. The hatched regions in the bottom two plots are a reminder: the FSA's predictions for observables in the delocalized phase are not meaningful. (a) The probability of a sample containing a resonance. The critical parameters are $\epsilon_c = -0.32(1)$ and $\tilde{\nu} = 0.75(6)$. (b) The disorder-averaged IPR within the FSA. The critical parameters are $\epsilon_c = -0.31(1)$ and $\tilde{\nu} = 0.83(7)$. (c) The disorder-averaged single-spin magnetization of the FSA wavefunctions. The critical parameters are $\epsilon_c = -0.30(1)$ and $\tilde{\nu} = 0.80(9)$.}
\label{fig:FSA_summary}
\end{center}
\end{figure}

To quantify resonance proliferation, we generate $\sim 10^3 - 10^5$ sets of REM energies and evaluate Eq.~\eqref{eq:FSA_expression}. First, we directly compute what we attempted to estimate in Sec.~\ref{subsec:single_resonance_approximation}: the probability of a sample at fixed $\Gamma$ and $\epsilon$ containing at least one resonance. In addition, we calculate the sample-averaged IPR and single-spin magnetization, for comparison with the exact diagonalization. These statistics each provide an independent measure of when resonances show up and how significant the resonances are. See Fig.~\ref{fig:FSA_summary} for scaled results, and refer back to Section~\ref{sec:exact_diagonalization} for definitions and notation.

To begin, we count the fraction $P$ of samples that contain a resonance at each value of $\Gamma$ and $\epsilon$. There is a crossover from $P \approx 0$ at large $|\epsilon|$ to $P \approx 1$ at small $|\epsilon|$ (Fig.~\ref{fig:FSA_summary}a). This crossover sharpens as $N$ increases, consistent with a zero-one law in the thermodynamic limit. We estimate the critical $\epsilon_c$ by finite-size scaling with the form $P = f(N^{\frac{1}{\tilde{\nu}}} (\epsilon - \epsilon_c))$. The curves collapse well, and $\epsilon_c$ is consistent with the exact diagonalization results. $\tilde{\nu}$, however, is significantly larger than ED predicts. The curve in Fig.~\ref{fig:FSA_summary}a has $\tilde{\nu} = 0.75(6)$. 

We next compute the disorder-averaged IPR (Fig.~\ref{fig:FSA_summary}b). For $\epsilon < \epsilon_c$, $[Y_2]$ increases towards $1$ as $N$ increases. 
This is consistent with localization and with the exact diagonalization results. It also confirms that only a negligible fraction of samples in this phase contain resonances. For $\epsilon > \epsilon_c$, however, $[Y_2]$ is roughly independent of both $\epsilon$ and $N$, and stays at $[Y_2] \approx 0.4$. This signifies resonances, and presumably delocalization, although the signature is very different from that of the exact eigenstates. The probability distributions for the wavefunction amplitudes are very long-tailed. Thus in a given sample, a few amplitudes will be much larger than the rest. In the localized phase, these largest amplitudes still do not compare to those of the initial site. As $\epsilon$ approaches $\epsilon_c$, those large amplitudes approach $1$ and $[Y_2]$ decreases. Once $\epsilon$ passes $\epsilon_c$, the initial site no longer contributes to the IPR and $[Y_2]$ is set by the largest non-initial amplitudes. Increasing $\epsilon$ (and $N$) even further will make the wavefunction amplitudes larger, but it won't change that only a few sites dominate the IPR. Hence $[Y_2]$ is independent of both $\epsilon$ and $N$ once the perturbation theory breaks down.

\begin{figure}[tb]
\begin{center}
\includegraphics[width=1.0\columnwidth]{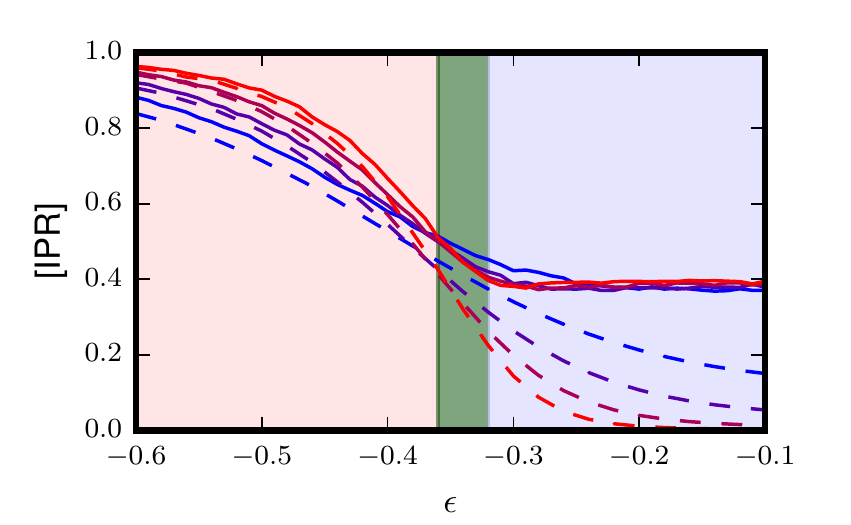}
\caption{A comparison of the disorder-averaged IPR  obtained from ED (dashed) and from the FSA (solid), at $\Gamma = 0.20$ and with $N$ ranging from $8$ (blue) to $14$ (red).
The background shading indicates the phase at that energy density as determined by the spectral statistics in ED: red is localized, blue is delocalized, and green is the transition region. Note that the FSA results are a good approximation in the localized region ($\epsilon < \epsilon_c$) but differ qualitatively in the delocalized region ($\epsilon > \epsilon _c$). This is not surprising, since the FSA is not valid in this region.}
\label{fig:FSA_ED_comparison}
\end{center}
\end{figure}

Keep in mind that the FSA's prediction for $[Y_2]$ in the delocalized phase has no connection to the actual IPRs in the QREM's delocalized phase. The FSA is a perturbative expansion, and all we learn from the initial site no longer contributing to $[Y_2]$ is that we cannot trust any results from the FSA in this region. Fig.~\ref{fig:FSA_ED_comparison} plots the FSA's disorder-averaged IPR against the disorder-averaged IPR from exact diagonalization. The two curves agree for all system sizes when $\epsilon < \epsilon_c$. The deviations become significant at $\epsilon = \epsilon_c$, though, and the curves bear no relation to each other for $\epsilon > \epsilon_c$. The FSA successfully describes the structure of eigenstates in the localized phase but not in the delocalized phase. Yet even though the FSA IPR is not accurate for much of the spectrum, a scaling collapse gives critical parameters that agree with those of the resonant-sample fraction $P$.

The magnetization of a single spin behaves analogously (Fig.~\ref{fig:FSA_summary}c). Recall that we arbitrarily take the unperturbed state to have all $\sigma_i^z = -1$, and sites closer to the initial site have more spins pointing down. Thus $[\avg{\sigma_1^z }]$ is roughly constant at $-1$ for large $|\epsilon|$, and it starts to increase significantly once $\epsilon$ passes $\epsilon_c$. As before, the FSA and ED predictions for $[\avg{\sigma_1^z }]$ agree in the localized phase but not in the delocalized phase. The curves collapse onto each other quite well, with scaling parameters that again agree with the others from the FSA.

\subsection{Replica Treatment} 
\label{sec:replica_treatment}

The statistical properties of the forward scattering wavefunctions can also be studied using the replica method, which provides complementary understanding to the other numerical and analytical approaches~\cite{Mezard:1987aa,Mezard:2009aa}. 
In this approach, we view the amplitude $\psi$ as the partition sum of a directed random polymer $p$ living on the hypercube with the long-tailed random weights $w_i = \Gamma/(E_a - E_i)$. 
As these weights do not have any finite moments, we expect the directed random polymer to condense onto a small number of large weight paths~\cite{kardar1994lectures}.
This also justifies our neglect of the sign of the weights $w_i$ as their destructive interference is unimportant in this regime.
The replica approach is especially useful as it naturally regulates the divergence of these moments.

Within the forward scattering approximation, the wavefunction amplitude $\psi_L$ at distance $L$ is
\begin{align}
	\psi_L &= \sum_{p}\prod_{i\in p} w_i \nonumber\\
	w_i &= \frac{\Gamma}{|E_0 - E_i|}
\end{align}
where $w_i$ is the random weight of the piece of path going through site $i$ on path $p$. 

The typical value of $\psi_L$ is provided by averaging its logarithm, $\overline{\ln \psi}$, which may be calculated using the formal replica trick
\begin{align}
	\overline{\ln \psi} = \lim_{n\to0} \frac{\overline{\psi^n} - 1}{n}
\end{align}
Thus, we need to calculate
\begin{align}
	\label{eq:replicamoments}
	\overline{\psi^n} &= \sum_{p_1,\cdots,p_n} \prod_{i} \overline{w_i^{r_i(p_1 \cdots p_n)}} 
\end{align}
where $r_i$ gives the number of times that the $n$ paths cross site $i$:
\begin{align}
	r_i&= \sum_{a=1}^n \mathbf{1}[i\in p_a]
\end{align}

\paragraph{Replica symmetric ansatz---} 
The replica symmetric ansatz consists of assuming that each of the $n$ paths contributes independently to the $n$'th moment of $\psi$. That is, $r_i = 1$ for each of the $nL$ sites visited by the paths and $0$ otherwise. Thus,
\begin{align}
	\overline{\psi^n}&\approx \sum_{p_1\cdots p_n} \prod_{i\in p_1\cdots p_n} \overline{w_i} = (L! \overline{w}^L)^n \\
	&= \exp\left[ n L (\ln L - 1 + \ln\overline{w})\right]
\end{align}
Thus, as $n \to 0$, we find
\begin{align}
	\overline{\ln \psi} = L (\ln L - 1 + \ln \overline{w})
\end{align}
This is ill-defined if $\overline{w} \to \infty$, which is clearly true for the weights arising in the QREM. However, for $E_0 = N \epsilon_0$, the weight in the tail of $w$ is parametrically suppressed by $N$. If we simply replace $\overline{w}$ by its typical value $\Gamma/N \epsilon_0$, we find the `typical weight RS' result,
\begin{align}
	\overline{\ln \psi} &\approx  L (\ln L - 1 + \ln\Gamma/N \epsilon_0) \nonumber \\
	&\approx  N l (\ln l  - 1 + \ln\Gamma/\epsilon_0)
\end{align}
where $L = l N$. 
This indicates that the typical amplitude decays exponentially with $N$ at all points inside the hypercube for $\Gamma < e \epsilon_0$. 

This clearly overestimates the critical $\Gamma$ for delocalization as it neglects the possibility of small denominators and the concomittant atypical resonances. 
Nonetheless, even at this level it shows that delocalization should take place for infinitesimal $\Gamma$ at $\epsilon_0 = 0$. 
This estimate agrees precisely with the rough estimate made directly in Sec.~\ref{subsec:rough_estimates}.

\paragraph{Replica symmetry breaking ansatz---}
In the 1RSB ansatz, the dominant configurations contributing to $\overline{\psi^m}$ consist of $n/x$ tightly bound groups of $x$ paths each. Thus,
\begin{align}
	\overline{\psi^n} &\approx \left(\sum_p \prod_{i\in p} \overline{w_i^x}\right)^{n/x} \nonumber \\
	&= \exp\left[ n f(x) \right]
\end{align}
where
\begin{align}
	f(x) & = \frac{L}{x} \left(\log L - 1 + \log \overline{w^x}\right)
\end{align}
is the 1RSB free energy function. In the $n\to 0$ limit, the Parisi parameter $x$ is constrained to the interval $[0,1]$ and the typical amplitude is given by the optimization
\begin{align}
\label{eq:rsblnpsi}
	\overline{\ln \psi} = \min_{x\in[0,1]} f(x)
\end{align}

The advantage of this approach is that it allows a direct treatment of the long-tailed weight distribution $p(w)$. Indeed, for $x<1$, the fractional moment $\overline{w^x}$ is convergent:
\begin{align}
	\overline{w^x} &= \int \frac{dE}{\sqrt{\pi N}} e^{-E^2/N} \left(\frac{\Gamma}{|E_0 - E|}\right)^x \nonumber\\
	&= \left(\frac{\Gamma}{\sqrt{N/2}}\right)^x \int \frac{du}{\sqrt{2\pi}} \frac{e^{-u^2/2} }{|u_0 - u|^x} \nonumber\\
	&= \left(\frac{\Gamma}{\sqrt{N/2}}\right)^x  I(x,u_0)
\end{align}
where $u_0 = E_0/\sqrt{N/2}$. The small denominator in the dimensionless integral $I(x,u_0)$ is integrable for $x<1$ and the Gaussian cuts off the power-law behavior at large $u$. 
Thus, for any fixed $u_0$ and $L$, $f(x)$ exhibits a positive divergence as $x\to 0^+$ and another as $x \to 1^-$, so that the minimizer $x^*$ lies strictly within the interval $(0,1)$ and all estimates are well-defined.

The 1RSB formalism is most useful at $E_0=0$, where the typical denominators are already quite small compared to the finite energy density case. For $u_0 = 0$, $I(x,0)$ reduces to a $\Gamma$ function:
\begin{align}
	I(x,0)  &= \frac{2^{-\frac{x}{2}} \Gamma \left(\frac{1-x}{2}\right)}{\sqrt{\pi }}
\end{align}
which exhibits the expected pole at $x=1$.

As $x\to 0$, $f(x)/L$ diverges as $(\log L - 1)/x$ while for $x \to 1$, $f(x)/L$ diverges with an $L$-independent logarithm. Thus, the minimizer $x^*$ must approach $1$ as $L$ grows, so we may simply replace $I(x,0)$ by its expansion near $x=1$ to linear order. In this approximation, we find that the saddle point of the replicated free energy arises at $x^* = 1- \frac{1}{\log \sqrt{2/\pi} n} + \cdots$ as $n\to\infty$, indicating condensation of the measure onto a logarithmically diverging subset of the paths.

Solving for the resonance condition $f = 0$ at $n=N$, we find the estimate
\begin{equation} 
\Gamma_c(\epsilon = 0) = \frac{\sqrt{\pi}}{2\sqrt{N}\log \sqrt{2/\pi}N} + \cdots
\end{equation}
for the critical value of the transverse field. We note that this estimate agrees to leading order in $1/\ln N$ with Eq.~\eqref{eq:high_T_critical_Gamma}.  

\section{Large-$\Gamma$ Limit} \label{sec:large_gamma_limit}

Here we describe the limit opposite to that of FSA, in which the random operator $E(\{\hat{\sigma}_i^z\})$ of Eq.~\eqref{eq:Hamiltonian} is the perturbation. This limit of $\Gamma \gg 1$ and the FSA limit of $\Gamma \ll 1$ are separated by a first-order thermodynamic transition, so we cannot extrapolate the FSA to high fields and must begin at $\Gamma \rightarrow \infty$.

To zeroth order in $E(\{\hat{\sigma}_i^z\})$, the eigenstates of the QREM Hamiltonian are $\{\sigma _i^x\}$ eigenstates with energies $-M\Gamma$ ($M = -N, -N+2, \ldots , N$). The degeneracy of the $-M\Gamma$ level is $\binom{N}{\frac{N+M}{2}}$. In this basis, the REM term is dense, i.e.,
\begin{equation} \label{eq:REM_x_basis}
\braket{\{\hat{\sigma}_i^x\}_a | E(\{\hat{\sigma}_i^z\})}{\{\hat{\sigma}_i^x\}_b} = \frac{1}{2^N} \sum _{j=1} ^ {2^N} (-1)^{\alpha _j ^{(a,b)}} E_j,
\end{equation}
where $a$ and $b$ denote different $\{\sigma _i^x\}$ eigenstates, $j$ enumerates the $\{\sigma _i^z\}$ eigenstates, and $\alpha _j ^{(a,b)}$ is 0 or 1 depending on the specific eigenstates in question. In words, every matrix element of $E(\{\hat{\sigma}_i^z\})$ is the sum and/or difference of the REM energies, divided by $2^N$ from normalization. Thus every matrix element is a Gaussian random variable of variance $\frac{1}{2^N} \frac{N}{2}$. We treat $E(\{\hat{\sigma}_i^z\})$ as a GOE-distributed random matrix in the $\{\sigma _i^x\}$ basis (this is only an approximation because the exact matrix elements aren't independent).

Under these conditions, the eigenstates within a subspace of fixed total magnetization $M$, which were initially degenerate, form a band according to Wigner's semi-circle law. This band has a half-width of
\begin{equation} \label{eq:x_band_radius_v1}
\sqrt{\frac{N}{2^{N+1}}} \sqrt{\binom{N}{\frac{N+M}{2}}}
\end{equation}
which in the large-$N$ limit is
\begin{equation} \label{eq:x_band_radius_v2}
\left( \frac{N}{2\pi (1- m^2)} \right) ^ {1/4} e^{N(\frac{1+m}{4}\ln{(1+m)} + \frac{1-m}{4}\ln{(1-m)})}
\end{equation}
where $m \equiv M/N$ is the average magnetization per spin.

All bands with $m \neq 0$ have widths much less than $\Gamma$ for large enough $N$. Thus the magnetization of each is preserved. Each \textit{single} spin does lose its magnetization because the eigenstates are statistically structureless within a band. Yet the only bands that hybridize are those with $M \ll N$.

Since the unperturbed energy densities are proportional to $m$, the large-$\Gamma$ situation is analogous to the small-$\Gamma$ situation: states at order-$N$ energies (finite temperature) are only weakly dressed by all others, while states at $\epsilon = 0$ (infinite temperature) hybridize immediately, i.e., regardless of how weak the perturbation is. However, it is difficult to carry out the large-$\Gamma$ expansion to higher orders, and so we cannot see the localization-delocalization transition from this side.

\section{Conclusion} 
\label{sec:conclusion}

The quantum random energy model, although non-local, is a useful test-bed for localization. Exact diagonalization shows a transition that appears to become sharp in the thermodynamic limit. This transition is visible through multiple order parameters. Furthermore, the forward scattering approximation to the perturbative-in-$\Gamma$ wavefunction agrees with ED results in the localized phase. Finite-size estimates of where perturbative resonances proliferate agree with the observed transition in ED as well.

Previous authors have explored replica-symmetry-breaking in the QREM's canonical ensemble. Our results suggest that this has little to do with the model's dynamics. This would not be all that surprising for a classical spin system since the classical Hamiltonian does not encode any dynamics. Yet a quantum spin's dynamics are fully determined by the Hamiltonian, and one would expect that its dynamics be compatible with its canonical ensemble. The QREM demonstrates otherwise.

Our finite-size numerics are consistent with a dynamical phase transition that is continuous. In particular, the autocorrelation time of spins diverges as the transition is approached from the delocalized phase. We can explain the variation of critical amplitudes with $\Gamma$ by a line of critical fixed points. However, our numerical observations are limited to small system sizes and the success of our scaling analysis may be finite-size effects.

Regardless, we find finite-size scaling windows controlled by the scaling combination $N \delta^{\tilde{\nu}}$. Exact diagonalization shows $\tilde{\nu} \approx 0.3-0.5$ and the numerical forward-scattering approximation shows a somewhat larger $\tilde{\nu} \approx 0.8$. Our $\tilde{\nu}$ should compared with $d \nu$ in $d$-dimensional disordered systems, which would be expected to satisfy the Harris criterion $d \nu \ge 2$ \cite{Chandran:2015ab,Harris:1974aa,PhysRevLett.57.2999}. 
The infinite-dimensional QREM need not and does not satisfy the Harris criterion, but it is consistent with observed $\nu$ in previous one-dimensional diagonalization studies \cite{kjall2014many,luitz2015many,Mondragon-Shem:2015aa}.

\section{Acknowledgements}

We would like to thank A.~Chandran, D.~Huse, V.~Oganesyan and G.~Parisi for discussions. CLB and CRL acknowledge the hospitality of ICTP where part of this work was completed. CLB acknowledges the support of the NSF through a Graduate Research Fellowship, Grant No. DGE-1256082. CRL would also like to acknowledge the hospitality of the Max-Planck Institute for the Physics of Complex Systems, Dresden and the support of the NSF through Grant No. PHY-1520535. Note that any opinion, findings, and conclusions or recommendations expressed in this material are those of the authors and do not necessarily reflect the views of the National Science Foundation.

\bibliography{biblioMBL}

\end{document}